\documentclass[10pt]{article}
\usepackage{fancyhdr}
\usepackage{extramarks}
\usepackage{amsmath}
\usepackage{amsthm}
\usepackage{amsfonts}
\usepackage{siunitx}
\usepackage{tikz}
\usepackage[plain]{algorithm}
\usepackage{algpseudocode}
\usepackage{multirow}
\usepackage{booktabs}
\usepackage{palatino}
\usepackage{graphicx}
\usepackage{subfigure}
\usepackage[colorlinks,linkcolor=black,anchorcolor=black,citecolor=black,urlcolor=blue]{hyperref}
\usepackage{amsmath,bm}
\usepackage{booktabs}
\usepackage{mathtools}
\usepackage{amssymb}
\usepackage{caption}
\usepackage{capt-of}
\usepackage{mciteplus}
\usepackage{cite}
\usepackage{mathrsfs}
\usepackage[title,titletoc,toc]{appendix}
\usepackage{xr}
\usepackage{parskip}
\usepackage{soul}
\usepackage{textcomp}
\usepackage[colaction]{multicol}
\usepackage[switch]{lineno}
\usepackage{lipsum}
\usepackage{etoolbox}
\usepackage{longtable}
\usepackage{array}
\usepackage{tablefootnote}
\newcolumntype{C}[1]{>{\centering\arraybackslash}p{#1}}
\usetikzlibrary{automata,positioning}
\usetikzlibrary{automata,positioning}

\begin{document}

\title{Review of COVID-19 Antibody Therapies}
% \author{Jiahui Chen\orcidA$^1$, Kaifu Gao\orcidB$^1$\footnote{Jiahui Chen and Kaifu Gao contributed equally.}, Rui Wang\orcidC$^1$, Duc Duy Nguyen\orcidD$^2$, and Guo-Wei Wei\orcidE$^{1,3,4}$\footnote{
% 		Corresponding author.		Email: wei@math.msu.edu} \\% Author name
% $^1$ Department of Mathematics, \\
% Michigan State University, MI 48824, USA.\\
% $^2$ Department of Mathematics, \\
% University of Kentucky, KY 40506, USA.\\
% $^3$ Department of Electrical and Computer Engineering,\\
% Michigan State University, MI 48824, USA. \\
% $^4$ Department of Biochemistry and Molecular Biology,\\
% Michigan State University, MI 48824, USA. \\
% }

\author{Jiahui Chen$^1$, Kaifu Gao$^1$\footnote{Jiahui Chen and Kaifu Gao contributed equally.}, Rui Wang$^1$, Duc Duy Nguyen$^2$, and Guo-Wei Wei$^{1,3,4}$\footnote{
		Corresponding author.		Email: wei@math.msu.edu} \\% Author name
$^1$ Department of Mathematics, \\
Michigan State University, MI 48824, USA.\\
$^2$ Department of Mathematics, \\
University of Kentucky, KY 40506, USA.\\
$^3$ Department of Electrical and Computer Engineering,\\
Michigan State University, MI 48824, USA. \\
$^4$ Department of Biochemistry and Molecular Biology,\\
Michigan State University, MI 48824, USA. \\
}
\date{\today} % Date for the report

\maketitle

\begin{abstract}
Under the global health emergency caused by coronavirus disease 2019 (COVID-19), efficient and specific therapies are urgently needed. Compared with traditional small-molecular drugs,  antibody therapies are relatively easy to develop and as specific as vaccines in targeting severe acute respiratory syndrome coronavirus 2 (SARS-CoV-2), and thus attract much attention in the past few months. This work reviews seven existing antibodies for SARS-CoV-2 spike (S) protein with three-dimensional (3D) structures deposited in the Protein Data Bank. Five antibody structures associated with SARS-CoV are evaluated for their potential in neutralizing SARS-CoV-2. The interactions of these antibodies with the S protein receptor-binding domain (RBD) are compared with those of angiotensin-converting enzyme 2 (ACE2) and RBD complexes. 
Due to the orders of magnitude in the discrepancies of experimental binding affinities, we introduce topological data analysis (TDA), a variety of network models, and deep learning to analyze the binding strength and therapeutic potential  of the aforementioned fourteen antibody-antigen complexes.  The current COVID-19 antibody clinical trials, which are not limited to the S protein target, are also reviewed.  
%In this work, the binding affinities between antibodies and S protein are summarized. However, these reported values are inconsistent in different literature due to various experimental conditions and techniques, which makes the potency of each antibody unclear. Therefore, we employ the TopNet model and graph network analysis to provide a unified model for the potency evaluation of SARS-CoV-2 S protein antibodies. Moreover, we collect the clinical antibody trails for COVID-19 worldwide, which do not specifically take S protein as the target. In conclusion, we give a review of potential antibody therapies for COVID-19 and provide a theoretical approach for the potency ranking of SARS-CoV-2 S protein antibodies.

\end{abstract}

\pagenumbering{roman}
\begin{verbatim}
\end{verbatim}

% {\setcounter{tocdepth}{4} \tableofcontents}
%
  \newpage
 %\clearpage
 %\pagebreak

% {\setcounter{tocdepth}{4} \tableofcontents}

\setcounter{page}{1}
\renewcommand{\thepage}{{\arabic{page}}}

% \begin{multicols}{2}
% \multicollinenumbers
% \linenumbers
\section*{Introduction}
\addcontentsline{toc}{section}{Introduction}
Coronavirus disease 2019 (COVID-19) pandemic caused by severe acute respiratory syndrome coronavirus 2 (SARS-CoV-2) %Severe acute respiratory syndrome coronavirus 2 (SARS-CoV-2) outbreak in Wuhan, China, in late December 2019 and 
has rapidly spread around the world. By June 17, 2020, more than 8.2 million individuals were infected and 443,000 fatalities had been reported. Currently, there are neither specific drugs nor effective vaccines available \cite{cao2020potent}. Traditional drug discovery involves a long and costly process, requiring more than 10 years to put an average drug on the market. Vaccine development typically takes  more than one year. In contrast, developing potent SARS-CoV-2 specified antibodies that are produced from blood B cells in response to and counteracting SARS-CoV-2 antigens is a less time-consuming and more efficient strategy for combating the ongoing pandemic \cite{wrapp2020structural, tian2020potent, wan2020human, hanke2020alpaca, ju2020human, cao2020potent, pinto2020structural, walls2019unexpected, wu2020noncompeting, yuan2020highly,rogers2020rapid}.

Antibody (Ab), also called immunoglobulin (Ig), is a large, Y-shaped protein that typically consists of two identical heavy chains  and two identical light chains. A heavy chain can be separated into two regions, the constant region and the variable region. Moreover, each light chain has two successive domains, the constant domain and the variable domain. The two heavy and two light chains of an antibody are connected through disulfide bonds within the constant region  \cite{putnam1979primary}.
 An antibody binds the antigenic determinant (also called epitope) through the variable regions in the tips of the heavy and light chains. Each of these chains contains three complementarity determining regions (CDRs), which located in the tips of each variable domain. Most of the diversities between antibodies are generated within the CDRs, which determine the specificity of individuals of antibodies. 

Benefit from the broad specificity of antibodies; antibody therapies have been proven to be a promising way to fight against cancer, autoimmune disease, neurological disorders, and infectious viruses such as HIV, Ebola, and Middle Eastern respiratory syndrome (MERS) \cite{corti2016protective, wang2018importance, scheid2009broad}. Recently, several studies have shown that the convalescent  plasma of SARS-CoV-2 patients, which contains neutralizing antibodies created by the adaptive immune response, can effectively improve patient survival rate \cite{chen2020convalescent, cao2020covid, shen2020treatment}. However, plasma-based therapeutics cannot be produced on a large scale. Therefore, seeking potent industrial-scale antibody therapies becomes one of the most feasible strategies to fight against SARS-CoV-2. The spike (S) protein, a multi-functional molecular machine that binds the human cell receptor angiotensin-converting enzyme 2 (ACE2), has been taking as the target of neutralizing antibodies and the focus of therapeutic and vaccine design efforts \cite{TORTORICI201993}. Many researchers  have reported the binding affinities of SARS-CoV-2 S protein in complex with antibody candidates and ACE2.  However, these reported values may vary by two orders of magnitude for a given antibody due to different experimental methods, conditions, calibrations, and human errors, which hinders the development of antibody therapies for SARS-CoV-2. Therefore, developing a unified paradigm for ranking the potency of SARS-CoV-2 antibodies is a pressing need. 

In this work, we review seven existing SARS-CoV-2 antibody therapeutic candidates  in the literature. As molecular structures are able to reveal the molecular mechanism of antibody-antigen interactions,  we only focus on the SARS-CoV-2 S protein antibodies that have three-dimensional (3D)  structures released on the Protein Data Bank (PDB). Since antibodies may directly compete with ACE2 for binding to the S protein receptor-binding domain (RBD), the structure and binding affinity of ACE2 and S protein complexes are studied to understand the efficiency of antibodies.   Moreover, since the S proteins of SARS-CoV and SARS-CoV-2 share 80\% amino acid sequence identity \cite{walls2020structure}, SARS-CoV S protein antibodies potential candidates for COVID-19. Therefore, we review five existing SARS-CoV S protein antibodies and analyze their binding affinities with the SARS-CoV-2 S protein.  Furthermore, we employ topological data analysis (TDA), artificial intelligence (AI), and a variety of network models to address literature controversy and provide a unified paradigm for ranking the potency of all antibodies. Finally,  we review the clinical trials of COVID-19 antibody candidates.

\section{An overview of antibody structures, functions, and therapies}
% \begin{figure}[ht]
%     \centering
%     \includegraphics[scale=0.40]{TOC_Antibody.pdf}
%     \caption{A schematic illustration of antibody therapy. }
%     \label{fig:AntibodyTherapy}
% \end{figure}

\begin{figure}[ht]
    \centering
    \includegraphics[scale=0.3]{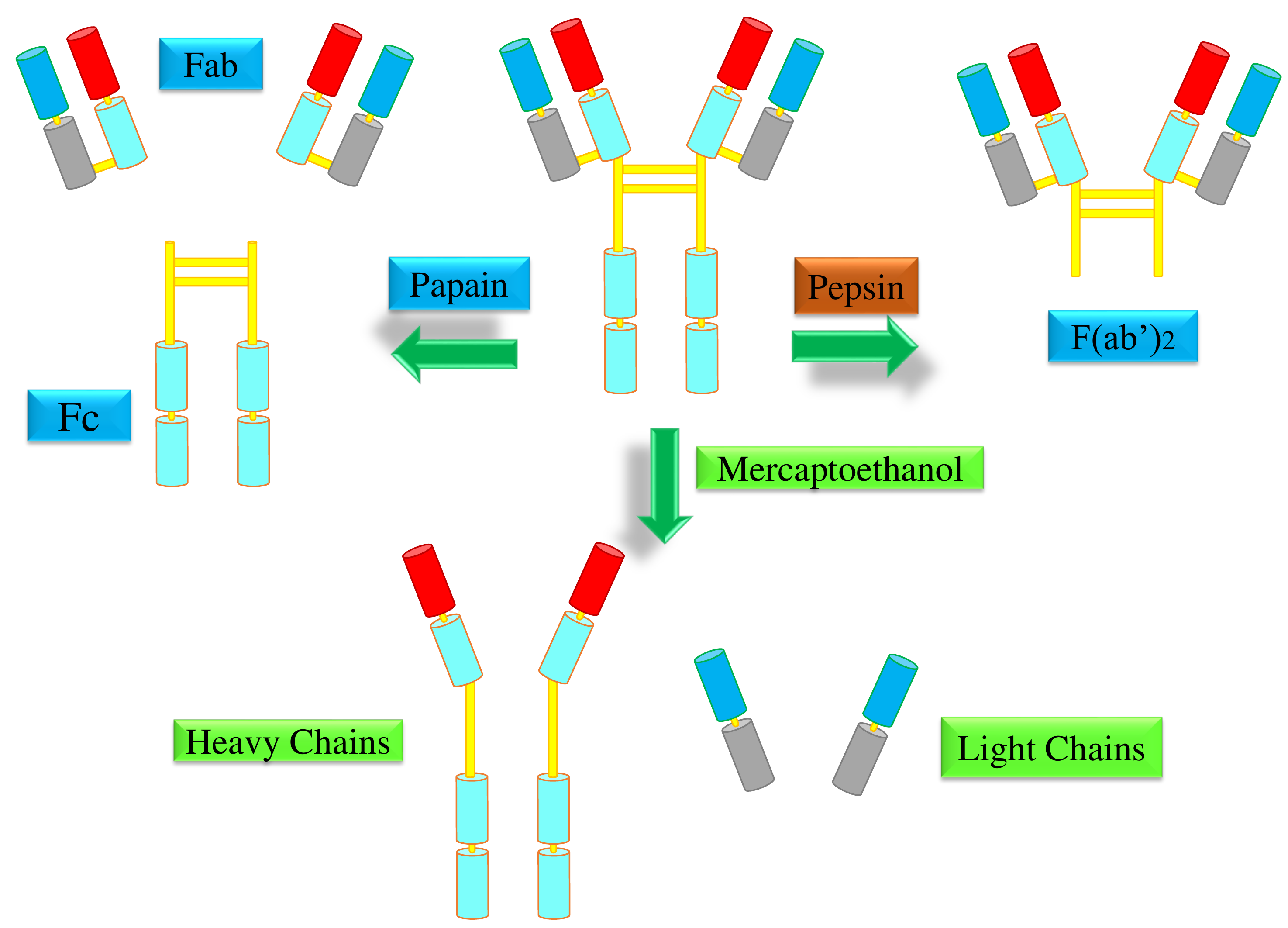}
    \caption{A schematic illustration  of antibody. }
    \label{fig:Antibody}
\end{figure}

An antibody can be divided into different parts according to its  functions. Specifically, the ``arms" of the Y-shaped protein contain sites that can recognize and bind to specific antigens. This region of the antibody is called a Fab (fragment, antigen-binding) region and composed of one constant domain and one variable domain from each heavy (VL) and light chain (VH) of the antibody \cite{putnam1979primary}. \autoref{fig:Antibody} illustrates the structure of the antibody. The variable domain (Fv region) is the most important region for binding to antigens. To be specific, on light and heavy chains, complementarity-determining regions (CDRs) composed of three variable loops of $\beta$-strands  are responsible for binding to a specific antigen. The CDRs are incredibly variable, allowing a large number of antibodies with slightly different tip structures, or antigen-binding sites, to exist. Each of these variants can bind to a different antigen so that the enormous diversity of antibody paratopes on the antigen-binding fragments allows the immune system to recognize an equally wide variety of antigens \cite{mian1991structure}. This antibody paratope diversity is generated by random recombination events of a set of gene segments that encode different antigen-binding sites (or paratopes), followed by random mutations in this area of the antibody gene to create further diversity \cite{market2003v, diaz2002somatic}. It has been estimated that humans generate about 10 billion different antibodies \cite{fanning1996development}. The base of the Y plays a role in modulating immune cell activity. This region is named an Fc (Fragment, crystallizable) region and is composed of two heavy chains. The Fc region ensures each antibody generates an appropriate immune response for a given antigen, by binding to a specific class of Fc receptors or other immune molecules. This process activates different physiological effects, including recognition of opsonized particles, lysis of cells, and degranulation of mast cells, basophils, and eosinophils \cite{woof2004human, heyman1996complement}.

In addition to conventional antibodies, camelids also produce heavy-chain-only antibodies (HCAbs). HCAbs, also referred to as VHHs, or nanobodies, contain a single variable domain (VHH) that makes up the equivalent antigen-binding fragment (Fab) of conventional immunoglobulin G (IgG) antibodies \cite{hamers1993naturally}. This single variable domain typically can acquire affinity and specificity for antigens comparable to conventional antibodies. VHHs can easily be constructed into multivalent formats and have higher thermal stability and chemostability than most antibodies do \cite{de2019single, dumoulin2002single, govaert2012dual, laursen2018universal, rotman2015fusion, van1999comparison}. Another advantage of VHHs is that they are less susceptible to steric hindrances than larger conventional antibodies \cite{forsman2008llama}. 

In immune systems, antibodies are generated and secreted by B cells, mostly differentiated B cells, including plasma cells or memory B cells. Antibodies have two physical forms, a membrane-bound form called the B-cell receptor (BCR), which is found to  attach to the surface of B cells, and a soluble form that moves freely in the blood plasma. The BCR facilitates the activation and subsequent differentiation of B cells into either plasma or memory B cells. The activation of B cells has two mechanisms: T cell-dependent (TD) activation and T cell-independent (TI) activation \cite{murphy2016janeway}. In the TD activation, once a BCR binds a TD antigen, the antigen is taken up into the B cell through receptor-mediated endocytosis, degraded, and presented to T helper (TH) cells as peptide pieces in complex with major histocompatibility complex-II (MHC-II) molecules on the cell membrane \cite{blum2013pathways}. TH cells recognize and bind these MHC-II-peptide complexes through their T cell receptor (TCR). Following TCR-MHC-II-peptide binding, T cells express the surface protein CD40L as well as cytokines such as IL-4 and IL-21. These signals promote B cell proliferation, immunoglobulin class switching, somatic hypermutation, and guide differentiation. Once receiving these signals, B cells are activated \cite{crotty2015brief}. In the TI activation, T cells are absent and B cells receive signals from recognition and binding of a common microbial constituent to toll-like receptors (TLRs) or extensive crosslinking of BCRs to repeated epitopes on a bacterial cell \cite{murphy2016janeway}. The TI activation is rapid, but antibodies generated from it tend to have a lower affinity and are also less functionally versatile than those from TD activation\cite{murphy2016janeway}. After activated, B cells can be differentiated into plasma cells or memory B cells to generate and secrete antibodies. Memory B cells can even survive in a human body for years to remember the same antigen and trigger a fast response upon future exposure \cite{borghesi2006b}. 

Antibodies protect our health in four ways: first, their Fab regions can bind to pathogens, so that prevent pathogens from entering or damaging cells; second, they trigger the removal of pathogens by macrophages and other cells by coating the pathogen; third, they stimulate the destruction of pathogens by stimulating other immune responses such as the complement pathway\cite{ravetch2001igg}; at last, antibodies can also lead to vasoactive amine degranulation against certain types of antigens such as helminths and allergens \cite{heyman1996complement}. 

The antibody mechanism enlightens the development of vaccines and antibody therapies. A vaccine is typically made of weakened or killed forms of the microbe, its toxins, or one of its surface proteins that resemble a disease-causing microorganism. These forms cannot cause an infection, but the immune system still regards them as a foreign object  and produces antibodies in response. After the threat has passed, most of the antibodies will break down, but memory B cells remain and remember the antigens in the vaccine. 

Antibody therapies were developed in the 1970s, following the discovery of the structures of antibodies and the development of hybridoma technology, which provided the first reliable source of monoclonal antibodies (mAbs) \cite{kohler1975continuous,breedveld2000therapeutic}. Rather than extracted from convalescent patient plasma, mAbs are  made from identical immune cells that are all clones of a unique parent cell so that they can have a monovalent affinity to the same epitope. As a result, the most significant advantage of  mAbs over conventional small-molecular drugs is their high specificity, which facilitates precise action \cite{hansel2010safety}. The second advantage is their long half-lives, which allows infrequent dosing \cite{leader2008protein}. Third, molecular engineering technologies have enabled the structure of mAbs to be fine-tuned for specific therapeutic actions and to minimize immunogenicity \cite{waldmann2005campath, nissim2008historical, presta2008molecular, hale2006therapeutic}, thus enhancing their safety.  This is reflected in mAbs having an approval rate of around 20\% compared with 5\% for new small-molecular entities \cite{reichert2005monoclonal, reichert2006anti}. Finally, mAbs can be developed  in a short time period, such as  5-6 months \cite{kelley2020developing}.  Currently, mAbs have already established their therapeutic and prophylactic efficacy against cancer, autoimmune  disease, neurological disorders, and infectious viruses such as HIV, Ebola, and Middle Eastern respiratory syndrome (MERS) \cite{corti2016protective, wang2018importance, scheid2009broad}. However, adverse effects are mostly relating to immunomodulation and infection can be associated with therapeutic mAbs \cite{hansel2010safety}, such as antibody-dependent enhancement \cite{tirado2003antibody} and cytokine storm \cite{clark2007advent}.

\section{SARS-CoV-2 antibody therapeutic candidates}

Both SARS-CoV and SARS-CoV-2  belong to lineage B of the betacoronavirus genus and have four structural proteins, known as  spike, envelope, membrane, and nucleocapsid proteins \cite{zhou2020pneumonia, wu2020new}. The nucleocapsid protein holds the RNA genome. Together with membrane, spike, envelope, and membrane proteins create the viral envelope \cite{wu2020analysis}. Among them, the S protein that forms homotrimers protruding from the viral surface mediates the entry of coronaviruses into host cells when binds with ACE2 \cite{TORTORICI201993}. More specifically, the S protein comprises two functional subunits: the S1  that is responsible for binding to the host cell receptor and the S2 that promotes the fusion of the virus and cellular membranes \cite{walls2016cryo,walls2017tectonic}. %The host cell receptor binding  prompts the S2 to transit from a metastable pre-fusion to a more-stable post-fusion state is essential for membrane fusion \cite{gui2017cryo, song2018cryo,kirchdoerfer2018stabilized,yuan2017cryo}. The S1 can further be divided into A, B, C, and D domains. Here, B domain is also called the RBD \cite{lan2020structure}, which binds to the angiotensin converting enzyme 2 (ACE2) on human cells.

ACE2 is a single-pass transmembrane protein with its active domain exposed on the cell surface and is expressed in the lungs and other tissues \cite{hamming2004tissue}. ACE2 serves as the main cell entry point for SARS-CoV and SARS-CoV-2, and some other coronaviruses \cite{walls2019unexpected}.  Notably, the  equilibrium dissociation constant (Kd) of the binding between ACE2 and S protein has significantly increased from SARS-CoV to SARS-CoV-2 \cite{walls2020structure,tian2020potent}. Moreover, plus SARS-CoV-2 may also use basigin to assist in cell entry \cite{wang2020sars}. Therefore, SARS-CoV-2 is more infectious than SARS-CoV.

\begin{figure}[ht]
    \centering
    \includegraphics[scale=0.5]{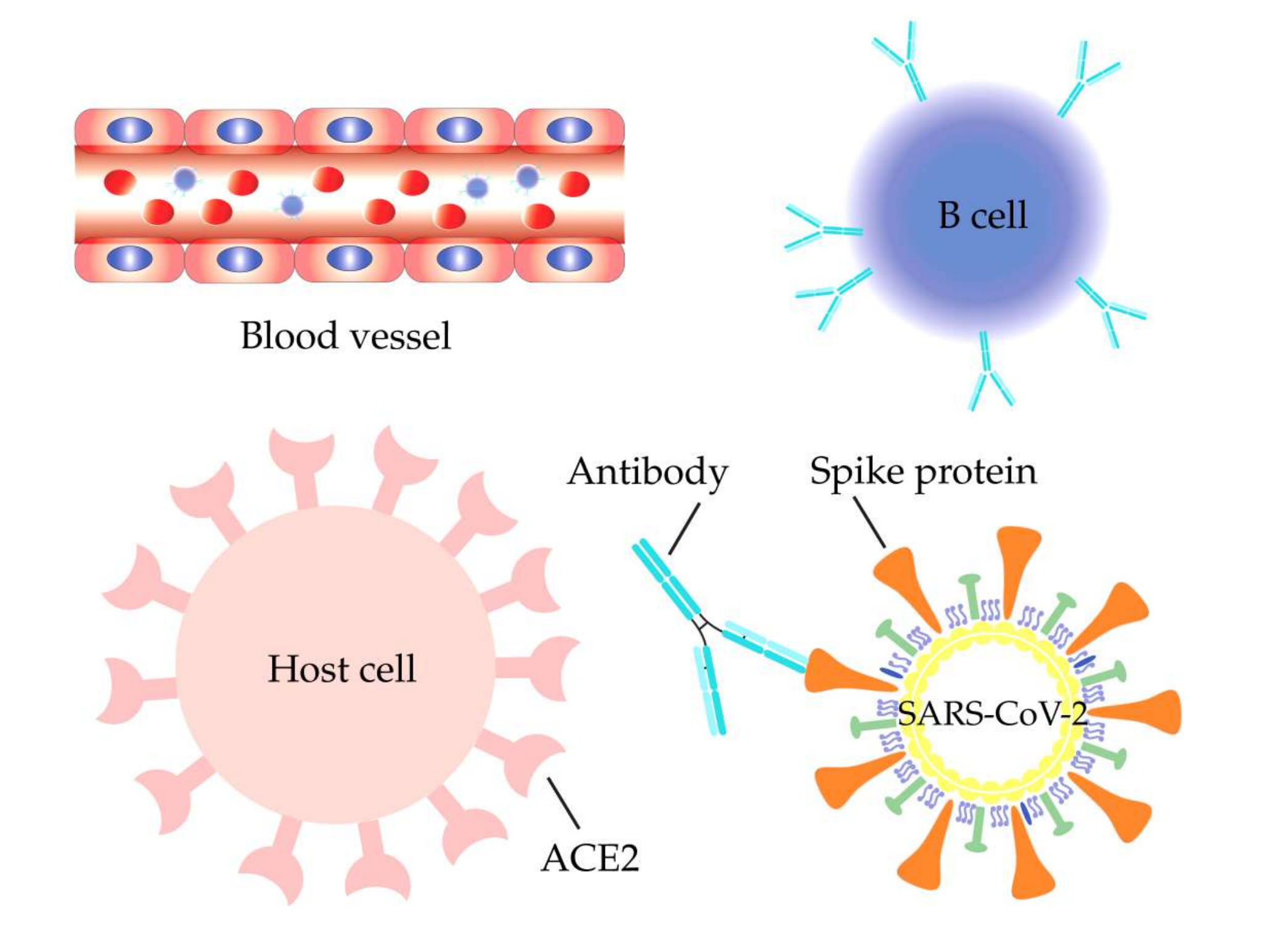}
    \caption{A schematic illustration of antibody therapy. }
    \label{fig:AntibodyTherapy}
\end{figure}

Antibody therapy is promising to fight against COVID-19. \autoref{fig:AntibodyTherapy} is a schematic illustration of antibody therapy for COVID-19.
%Recent reports have shown that the survival rate of SARS-CoV-2 patients' improved by using the convalescent patients’ plasma \cite{chen2020convalescent, cao2020covid, shen2020treatment}. However, such plasma-based therapeutic is quite limited since plasma cannot be produced broadly. 
Notably, neutralizing monoclonal antibodies (mAbs) isolated from convalescent patient memory B cells provide effective intervention to SARS-CoV-2 due to their safety, scalability, and therapeutic effectiveness \cite{chen2020convalescent, cao2020covid, shen2020treatment}. As  S protein mediates the  host cell entry, it is the target of neutralizing antibodies and the focus of therapeutic and vaccine design efforts \cite{TORTORICI201993}. %Despite their advantages, screening for potent neutralizing mAbs from human memory B cells is often a slow and laborious process, which is not ideal when responding to a worldwide health emergency. A rapid and efficient method for screening SARS-CoV-2-neutralizing mAbs is urgently needed (7byr) %The S trimers are extensively decorated with N-linked glycans that are important for protein folding11 and modulate accessibility to host proteases and neutralizing antibodies12-15. 

%which target  the SARS-CoV-2 spike (S) protein   receptor-binding domain  (RBD) of  angiotensin-converting enzyme 2 (ACE2). They are S309, CR3022,CB6, P2B-2F6, H11-D4. Moreover, ACE2 is also taken into consideration for the comparison analysis with the SARS-CoV-2-antibodies mentioned above. Moreover, {\color{red} eight} antibodies that target to SARS-CoV RBD are studied as well, which are listed in \autoref{table:Summary1}.

\begin{table}[ht]
	\centering
	\setlength\tabcolsep{1pt}
	\captionsetup{margin=0.9cm}
	\caption{A summary of monoclonal antibodies targeting the SARS-CoV-2 or SARS-CoV S-protein with the 3D experimental structures of their complexes available in Protein Data Bank. BLI: Biolayer interferometry, and SPR: Surface plasmon resonance.}
	\begin{tabular}{l|l|l|l|c}       
		\hline
		Protein or Antibody & Target & Kd (nM) / Method  & PDBID &Resolution (\AA)  \\ \hline
		&  & 1.2 / BLI \cite{walls2020structure}  &  &     \\ 	
		ACE2 & SARS-CoV-2 RBD &  34.6 / BLI\cite{wrapp2020cryo}    & 6M0J \cite{lan2020structure} & 2.45  \\ 	
		&  &  15.2 / BLI \cite{tian2020potent}  & &    \\ \hline	
		S309 & SARS-CoV-2 RBD & IgG=0.104 Fab=1.98 / BLI \cite{pinto2020structural}  & 6WPS, 6WPT \cite{pinto2020structural} & 3.10, 3.70    \\ \hline
		CR3022 & SARS-CoV-2 RBD & IgG=6.28 / BLI \cite{tian2020potent}  & 6W41 \cite{yuan2020highly} &  3.08  \\ 
		&  & IgG\textless0.1 Fab=115  / BLI \cite{yuan2020highly} &  &   \\ \hline
		% 		& /& IgG\textless0.1 Fab=115 \cite{yuan2020highly} & & \\ 	
		CB6 &  SARS-CoV-2 RBD & IgG=2.49 / SPR \cite{shi2020human}  &7C01  &  2.85  \\\hline	
		P2B-2F6  & SARS-CoV-2 RBD & IgG=5.14 / SPR \cite{ju2020human}  & 7BWJ \cite{ju2020human} & 2.85 \\		\hline		
%		VHH-72 & SARS-CoV-2 RBD & IgG=38.6 / SPR \cite{wrapp2020structural} & &  \\ \hline		
		B38 &  SARS-CoV-2 RBD & IgG=70.1 / SPR\cite{wu2020noncompeting} & 7BZ5 \cite{wu2020noncompeting} &  1.84  \\ \hline
		%		Ty1 &  SARS-CoV-2 S-protein & Fab=54 \cite{hanke2020alpaca}& & \\ 
		H11-D4 & SARS-CoV-2 RBD &  & 6Z43 &3.30  \\	\hline
		BD23 & SARS-CoV-2 RBD &  & 7BYR\cite{cao2020potent} &3.84  \\	\hline
		&  & 5.0 / BLI \cite{walls2020structure}   &   & \\ 	
		ACE2 & SARS-CoV RBD & 325.8 / BLI \cite{wrapp2020cryo}  & 3D0G & 2.80  \\ 
		&  & 1.70 / BLI  \cite{sui2004potent} &   &   \\ \hline
		CR3022 & SARS-CoV RBD & IgG\textless0.1 Fab=1 / BLI \cite{yuan2020highly} &  &   \\ \hline
		S309 & SARS-CoV RBD & IgG=0.12 Fab=1.81 / BLI \cite{pinto2020structural} &  &  \\ \hline
		m396 & SARS-CoV RBD & IgG=0.005 Fab=20 / BLI \cite{prabakaran2006structure}  & 2DD8 \cite{prabakaran2006structure} & 2.30  \\ \hline
		S230 & SARS-CoV RBD & IgG=0.06 / BLI \cite{walls2019unexpected}  & 6NB6 \cite{walls2019unexpected} & 4.30  \\ 		\hline
		VHH-72\footnotemark & SARS-CoV RBD & IgG=1.15 / SPR \cite{wrapp2020structural}  & 6WAQ \cite{wrapp2020structural} &  2.20 \\ 	\hline				
		80R & SARS-CoV RBD & IgG=1.59 / BLI \cite{sui2004potent}  & 2GHW \cite{hwang2006structural} &  2.30 \\ \hline
		F26G19 &  SARS-CoV RBD & Fab=26 / SPR \cite{pak2009structural}  & 3BGF \cite{pak2009structural} &  3.00 \\ \hline	
	\end{tabular}
	\label{table:Summary1}
\end{table}

Table \ref{table:Summary1} provides a summary of  SARS-CoV-2 and SARS-CoV S protein RBDs in complexes with existing  antibodies and ACE2 structures.  The structures, functions, and properties of these complexes are analyzed below. 

As summarized in Table \ref{table:Summary1}, twelve mAbs targeting the SARS-CoV-2 or SARS-CoV S-protein RBD are reported with their 3D experimental complex structures  available in the Protein Data Bank. According to the literature reports, the most promising one is S309 \cite{pinto2020structural}, which shows almost equally neutralization potency against both SARS-CoV and SARS-CoV-2. The authors state that 19 out of 24 residues of the S309's epitope are strictly unchanged from SARS-CoV and SARS-CoV-2, and the other 5 residues are conservatively or semi-conservatively substituted \cite{pinto2020structural}. However, some other researchers are still concerned about the claimed cross-effectiveness against both SARS-CoV and SARS-CoV-2 \cite{wan2020human}. Notably, two experimental structures of the S309 and SARS-CoV-2 S-protein complex are released with  closed and open conformations of the S-protein, respectively. The binding affinity of S309 and S protein RBD complex is not sensitive to S protein close or open conformations \cite{pinto2020structural}.  

CR3022 is another potent antibody that may bind to both SARS-CoV and SARS-CoV-2 \cite{yuan2020highly, tian2020potent}. It is also showed that compared with m396, a SARS-CoV–specific antibody, CR3022 has a significantly stronger binding signal to SARS-CoV-2. However, this affinity to SARS-CoV-2 is much weaker than the affinity to SARS-CoV \cite{yuan2020highly}. It is also suggested that CR3022 can only access to the open conformation of the S-protein RBD \cite{yuan2020highly}.

CB6 and P2B-2F6 are also promising SARS-CoV-2 antibodies, which are specific human mAbs extracted from convalescent COVID-19 patients \cite{shi2020human, ju2020human}.
VHH-72 cross-reacts with SARS-CoV-2 and SARS-CoV S-proteins, but the affinity to SARS-CoV-2 is much lower than SARS-CoV \cite{wrapp2020structural}. B38 also shows a direct competition with ACE2 for binding to  SARS-CoV-2 S-protein \cite{wu2020noncompeting}.
\footnotetext{The binding affinity of VHH-72 with SARS-CoV-2 RBD is Fab$=54$ nM.}

\begin{figure}[ht]
	\centering
	\includegraphics[width=0.6\textwidth]{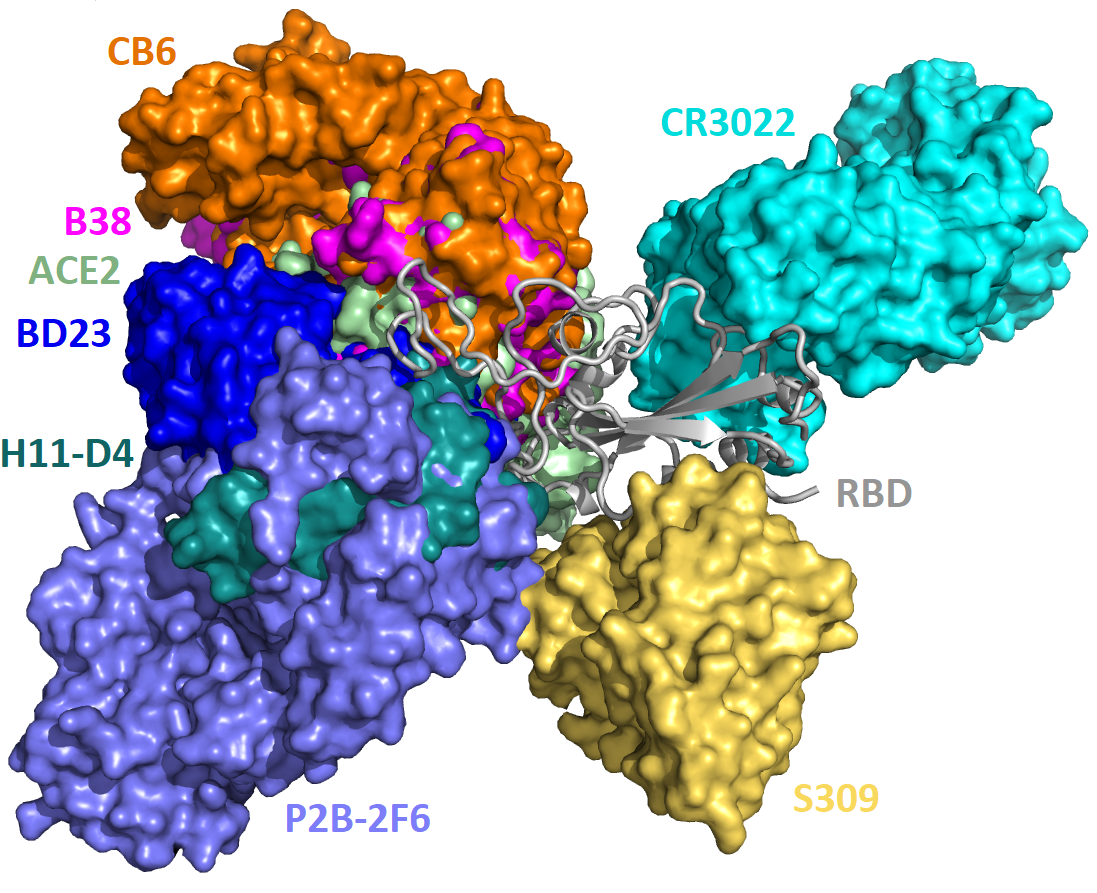}
	\caption{The alignment of the available 3D structures of  SARS-CoV-2 S-protein RBD in binding complexes with  antibodies as well as ACE2.}
	\label{fig:sars2-antibody-3d-alignment}
\end{figure}

\subsection{3D structure alignment}
\begin{figure}[ht]
	\centering
	\includegraphics[width=0.6\textwidth]{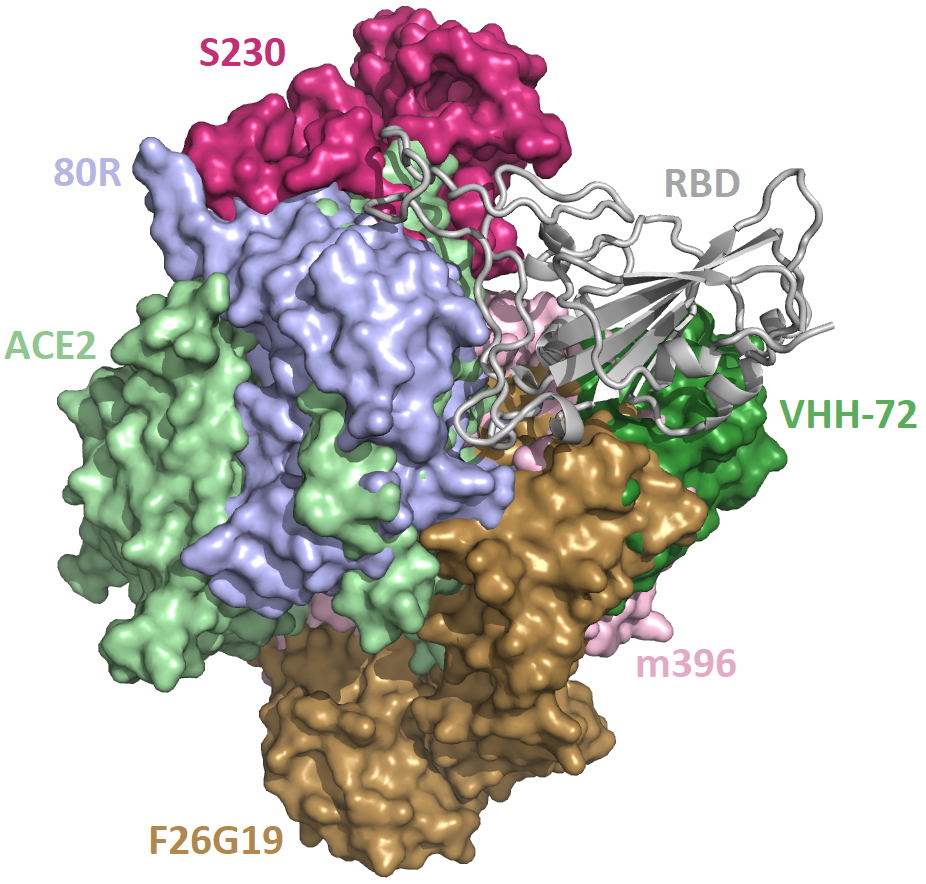}
	\caption{The alignment of the available 3D structures of  SARS-CoV S-protein RBD in binding complexes with  antibodies as well as ACE2.}
	\label{fig:sars-antibody-3d-alignment}
\end{figure}

All the available 3D structures of SARS-CoV-2 S protein RBD in complex of the antibodies are aligned to ACE2.  Figures \ref{fig:sars2-antibody-3d-alignment} and  \ref{fig:sars-antibody-3d-alignment} show the alignment of SARS-CoV-2 and SARS-CoV antibodies, respectively. The PDB ID of these complexes can be found in Table \ref{table:Summary1}.

As revealed in Figure \ref{fig:sars2-antibody-3d-alignment}, the antibodies CB6, B38, H11-D4, and P2B-2F6 have their epitopes overlapping with the ACE2 binding site, suggesting their bindings are in direct  competition with that of ACE2. Theoretically, this direct competition reduces the viral infection rate. For an antibody with strong binding ability, it will directly neutralize SARS-CoV-2 without  the need of  antibody-dependent cell cytotoxicity (ADCC) and antibody-dependent cellular phagocytosis (ADCP) mechanisms.  However, the binding sites of epitopes of S309 and CR3022 are away from that of ACE2, leading to the absence of binding competition \cite{pinto2020structural, tian2020potent}. One study shows the  ADCC and  ADCP mechanisms contribute to the viral control conducted by S309 in infected individuals \cite{pinto2020structural}. For CR3022, researches indicate that it neutralizes the virus in a synergistic fashion \cite{ter2006human}.

  Figure \ref{fig:sars2-antibody-3d-alignment} shows that on the SARS-CoV RBD, antibodies S230, 80R, F26G19, and m396 have their epitopes overlapping with ACE2.  
 VHH-72 locates slightly away from the ACE2 binding site but still sterically clashes with the ACE2 binding. They all lead to binding-competitions to naturalize the virus.

\subsection{Alignment of antibody and ACE2 epitopes on   S protein two-dimensional (2D) sequences }

\addcontentsline{toc}{subsection}{Alignment of antibody or ACE2 epitopes on   S protein two-dimensional (2D) sequences }

\begin{figure}[ht]
	\centering
	\includegraphics[width=1.0\textwidth]{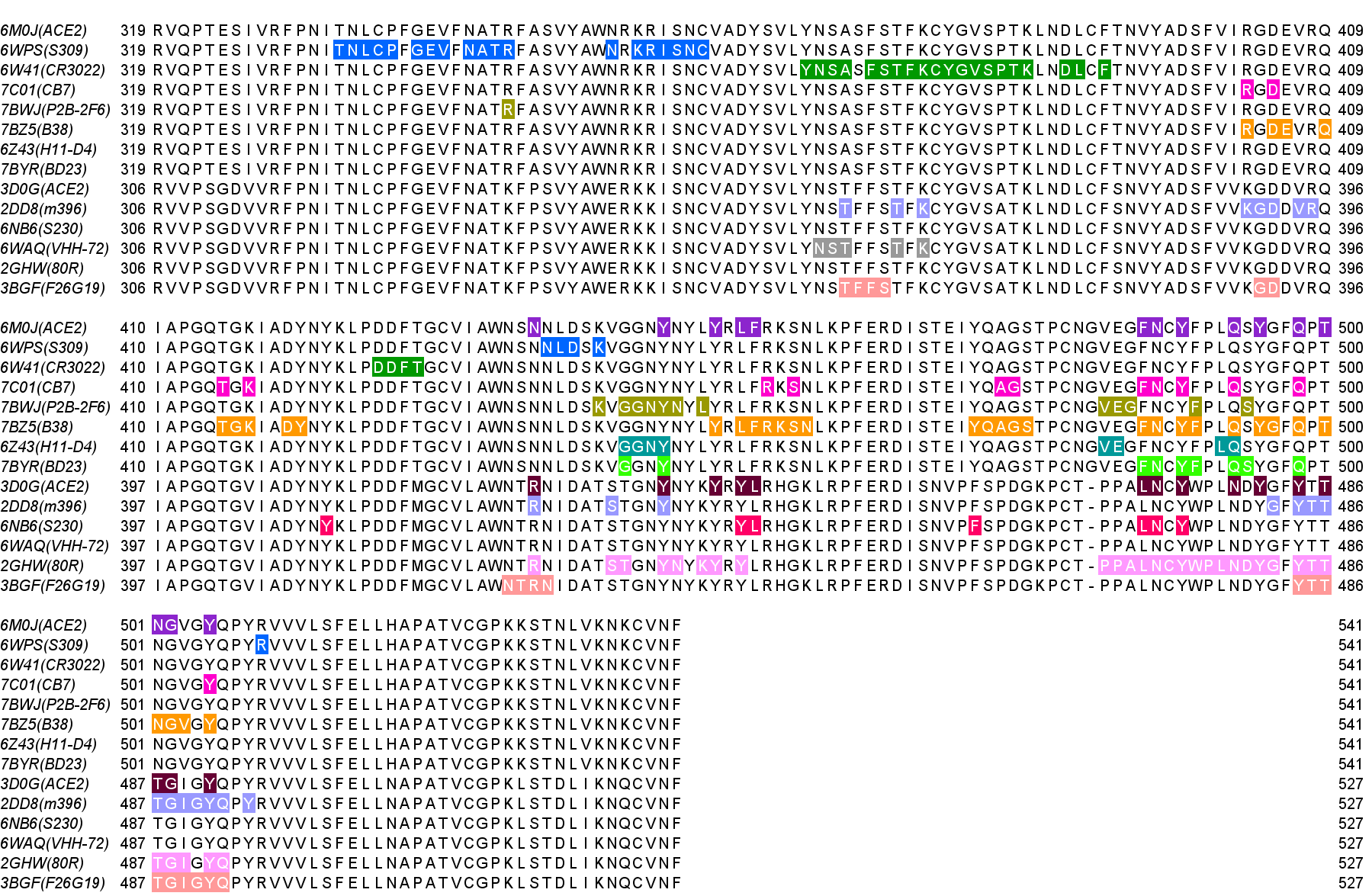}
	\caption{Illustration of the contact positions of  antibody and ACE2 epitope with SARS-CoV-2 and SARS-CoV S protein RBDs on RBD two-dimensional (2D) sequences. The proteins in the structures of 6M0J, 6WPS, 6W41, 7C01, 7BWJ, 7BZ5, 6Z43, and 7BYR are in complexes with  SARS-CoV-2 S-protein while the proteins in structures 3D0G, 2DD8, 6NB6, 6WAQ, 2GHW, and 3BGF are in complexes with SARS-CoV S protein.}
	\label{fig:sars-antibody-2d-align}
\end{figure}

Figure \ref{fig:sars-antibody-2d-align} highlights the contact regions of antibody and ACE2 epitopes on SARS-CoV-2 RBD or SARS-CoV RBD 2D sequences.  Consistent with the 3D alignment, except for S309, CR3022, and VHH-72, all the other antibodies have their epitopes overlapping with the ACE2 binding site, especially the residues from 486 to 505 of the SARS-CoV-2 RBD (corresponding to the residues 472 to 491 of the SARS-CoV RBD). Although the VHH-72 epitope residues do not overlap with the ACE2 binding site, VHH72 occupies parts of the space of the ACE2 binding site, which will disrupt  the ACE2 binding with RBD. Therefore,  VHH-72 also has a competitive binding ability against ACE2. Figure \ref{fig:sars-antibody-2d-align} also reveals that these epitope residues have many mutations from SARS-CoV-2 RBD to SARS-CoV RBD, which could explain why most of the antibodies lack cross-reaction to both SARS-CoV-2 and SARS-CoV. This aspect will be further studied in a later section. 

\section{Experimental pitfalls in the affinity measurements of antibody binding with S protein RBD}

 Table \ref{table:Summary1} clearly displays the discrepancy in reported experimental Kd values for ACE2 in complexes with SARS-CoV-2 S-protein (i.e., 1.2 nM \cite{walls2020structure}, 15.2  nM \cite{tian2020potent},  and  34.6 nM \cite{wrapp2020cryo}).  Moreover, 
a 191-fold difference in magnitude has been reported in experimental Kd values 
for ACE2 and  SARS-CoV S-protein complex (i.e., 5.0 nM \cite{walls2020structure}, 325.8 nM  \cite{wrapp2020cryo}, and 1.70 nM  \cite{sui2004potent}). 

The inconsistency mentioned above in experimental values is not isolated. The experimental Kd values for CR3022 binding with the SARS-CoV-2 S protein RBD were reported as  6.28 nM  \cite{tian2020potent}  and \textless0.1  nM  \cite{yuan2020highly}.  This level of  discrepancy in reported experimental values makes it impossible to select antibody candidates appropriately. 

As shown in Table \ref{table:Summary1},  two binding assay techniques are used to measure Kd values of antibody-antigen interactions.  The discrepancies in Kd values for CR3022 discussed above are based on 
biolayer interferometry (BLI) measurements.  BLI detects the surface changes on biosensor tips induced by protein-protein association and dissociation by using  analyzing the interference pattern of white light reflected from the surface. BLI results are sensitive to biosensor preparation, the stability of the  light source, the temperature control,  and the calibration.  Surface plasmon resonance (SPR) has been employed for determining  the Kd values of antibody and RBD complexes as shown in  Table \ref{table:Summary1}. This method detected the reflectivity changed  induced by molecular adsorption, such as polymers, DNA or proteins, etc. by changes in reflection angles. Similarly, SPR is also sensitive to the preparation of conjugated antigens, the stability of the  light source, the temperature control, and the calibration. The widespread inconsistency in reported antibody and S protein binding affinities motivates us to carry out computational analysis of existing antibody-S protein  complexes described below. 

\section{Computational analysis of antibody-SARS-CoV-2 interactions}
 
To have  unified assessment and ranking of S protein RBD binding complexes with antibodies and ACE2, we utilize topological data analysis, graph theory, network models, and machine learning to analyze the 3D complexes presented in Table \ref{table:Summary1}. We also evaluate the re-purposing potentials of SARS-CoV antibodies for SARS-CoV-2 using the TopNetTree model \cite{wang2020topology}. 

\subsection{Ranking of ACE2 and antibodies}

The prediction results and network descriptors are presented in Table~\ref{tab:SARS2} and Table~\ref{tab:SARS1} for SARS-CoV-2 complexes and SARS1 complexes, respectively. In Table~\ref{tab:SARS2}, the complexes are ranked according to their predicted binding affinities, $\Delta G$, followed by FRI rigidity indexes \cite{nguyen2016generalized, xia2013multiscale} which have the highest covariance. They are computed based on all the  C$_\alpha$ atoms on the RBD and all the  C$_\alpha$ atoms in antibodies or ACE2. 
The FRI rigidity index $R_{\eta}$ indicates the measurement of geometric compactness and topological connectivity of protein-protein interactions at each residue, such that more impact of the longer pairwise interactions for larger $\eta$. Comparing with the predicted energy, a strong binding affinity is corresponding to a large rigidity index. Then the summation of binding affinity changes computed with the TopNetTree model \cite{wang2020topology} by modifying the RBD residues to glycine (G) are presented while $S_{10}$ and $S_8$ stand for those residues within 10{\AA } and 8{\AA } to the binding site are included. Each mutation to glycine (G) is calculated by PPI machine learning model \cite{wang2020topology}, where a positive binding affinity changes $\Delta\Delta G$ means a stronger binding affinity of mutant and vice versa. Therefore, a summation of considered residues in the RBD with smaller values indicates a strong binding affinity of wild type. 

The rest of the table gives the C$_{\alpha}$-based complex analysis from a number of network descriptors, including 
edge density ($d$),  
degree heterogeneity ($\rho$) \cite{estrada2010quantifying},
average path length ($\langle L \rangle$) \cite{watts1998collective},
average betweenness centrality ($\langle C_b \rangle$)  \cite{freeman1978centrality},
average eigencentrality ($\langle C_e \rangle$) \cite{bonacich1987power}, 
average subgraph centrality ($\langle C_s \rangle$) \cite{estrada2005subgraph}, 
and  network communicability ($\langle M \rangle$)  \cite{estrada2008communicability}.
Except for the degree heterogeneity which is calculated based only on the RBD C$_\alpha$ atoms, other descriptors are calculated from all   C$_\alpha$ atoms on the RBD and  antibody (or ACE2) C$_\alpha$ atoms within 10{\AA } from any  C$_\alpha$ atom on the RBD. The degree heterogeneity demonstrates  antibody or ACE2 influence to the RBD, such that close degree heterogeneity numbers would have similar impacts. For example, molecules B38 and H11-D4 have degree heterogeneity values that are close to ACE2 as well as sharing the same receipt domain. As for other descriptors, the average betweenness centrality \cite{freeman1978centrality} and average eigencentrality \cite{bonacich1987power} values are correlated quite well to the predicted binding affinities.

\begin{table}[]
\centering
\caption{The graph network descriptor consisting of SARS-CoV-2 Spike protein RBD and ACE2 and antibodies. $\Delta G$: the predicted binding affinity (kcal/mol) (the predictions are using the Prodigy web server \cite{xue2016prodigy}); $\text{R}_{10}$ and $\text{R}_{8}$: FRI rigidity index with $\eta$ of 10 and 8, respectively; $\text{S}_{10}$ and $\text{S}_{8}$: the summation of binding affinity changes ($\Delta\Delta G$ kcal/mol) by changing RBD residues within 10{\AA } and 8{\AA } to the binding site to glycine; $d$: edge density; $\rho$: degree heterogeneity; $\langle L \rangle$: average path length; $\langle C_b \rangle$: average betweenness centrality; $\langle C_e \rangle$: average eigencentrality; $\langle C_s \rangle$: average subgraph centrality; $\langle M \rangle$: network communicability.
}
\begin{tabular}{@{}|c||r|r|r|r|r|r|r|r|}  \hline
\multicolumn{1}{|c||}{Molecule} & \multicolumn{1}{c|}{CR3022} & \multicolumn{1}{c|}{B38} & \multicolumn{1}{c|}{CB6} & \multicolumn{1}{c|}{ACE2} & \multicolumn{1}{c|}{BD23} & \multicolumn{1}{c|}{H11-D4} & \multicolumn{1}{c|}{S309} & \multicolumn{1}{c|}{P2B-2F6} \\ \hline
\multicolumn{1}{|c||}{PDB ID} &
\multicolumn{1}{c|}{6W41} & \multicolumn{1}{c|}{7BZ5} & \multicolumn{1}{c|}{7C01} & \multicolumn{1}{c|}{6M0J} & \multicolumn{1}{c|}{7BYR} & \multicolumn{1}{c|}{6Z43} & \multicolumn{1}{c|}{6WPS} & \multicolumn{1}{c|}{7BWJ} \\ \hline \hline
\multicolumn{1}{|c||}{$\Delta G$}            & -15.4 & -14.7 & -13.5 & -11.9 & -10.8 & -10.3 & -9.9 & -9.6 \\ \hline
\multicolumn{1}{|c||}{$\text{R}_{10}$}       & 335 & 349 & 338 & 279 & 227 & 201 & 256 & 211 \\ \hline
\multicolumn{1}{|c||}{$\text{R}_{8}$}        & 134 & 138 & 132 & 105 & 106 & 82 & 97 & 82 \\ \hline
\multicolumn{1}{|c||}{$\text{S}_{10}$}       & 19.15 & 30.17 & 36.56 & 20.83 & 10.39 & 8.91 & 18.28 & 17.74 \\ \hline
\multicolumn{1}{|c||}{$\text{S}_{8}$}        & 11.69 & 12.39 & 13.36 & 16.60 & 5.49 & 4.41 & 7.04 & 5.97 \\ \hline
\multicolumn{1}{|c||}{$d$}                   & 0.070 & 0.069 & 0.072 & 0.072 & 0.069 & 0.077 & 0.071 & 0.074 \\ \hline
\multicolumn{1}{|c||}{$\rho$}                & 0.0192 & 0.0190 & 0.0196 & 0.0185 & 0.0171 & 0.0190 & 0.0206 & 0.0196 \\ \hline
\multicolumn{1}{|c||}{$\langle L \rangle$}   & 13.69 & 14.26 & 13.75 & 13.85 & 14.80 & 13.59 & 13.86 & 13.98 \\ \hline
\multicolumn{1}{|c||}{$\langle C_b \rangle$} & 0.0109 & 0.0111 & 0.0110 & 0.0113 & 0.0130 & 0.0113 & 0.0112 & 0.0117 \\ \hline
\multicolumn{1}{|c||}{$\langle C_e \rangle$} & 0.052 & 0.050 & 0.052 & 0.051 & 0.053 & 0.054 & 0.054 & 0.053 \\ \hline
\multicolumn{1}{|c||}{$\langle C_s \rangle$} & 1590955 & 2397825 & 2010826 & 2105421 & 813061 & 2248985 & 1387110 & 1562243 \\ \hline
\multicolumn{1}{|c||}{$\langle M \rangle$}   & 847509 & 1237464 & 1096771 & 1132331 & 413572 & 1239452 & 753625 & 855641 \\ \hline 
\multicolumn{1}{|c||}{$\langle \Theta \rangle$} & 0.0192 & 0.0190 & 0.0196 & 0.0185 & 0.0171 & 0.0190 & 0.0206 & 0.0196 \\ \hline
\end{tabular}
\label{tab:SARS2}
\end{table}

Table~\ref{tab:SARS1} shows the results of predictions and network descriptors for the SARS-CoV S  protein complex. Again, the predicted binding affinities have high correlations to  FRI rigidity indexes. For degree heterogeneity, m396 has a similar impact as ACE2. In Table~\ref{tab:SARS1}, molecule 80R (PDB 2GHW) has the highest rigidity index both for $\eta=10$ and $\eta=8$ which indicates a more rigid complex structure between 80R and the RBD. In the comparison of SARS-CoV S protein RBD and SARS-CoV-2 S protein RBD in Table~\ref{tab:SARS2}, descriptors are close to each other except for the summation of binding affinity changes which includes more biological and chemical information than others. Thus, the network structures for SARS-CoV RBD and SARS-CoV-2 RBD complex are similar.

\begin{table}[]
\centering
\caption{The graph network descriptor consisting of SARS-CoV Spike protein RBD and ACE2 and antibodies. $\Delta G$: the predicted binding affinity (kcal/mol) (the predictions are using the Prodigy web server \cite{xue2016prodigy}); $\text{R}_{10}$ and $\text{R}_{8}$: FRI rigidity index with $\eta$ of 10 and 8, respectively; $\text{S}_{10}$ and $\text{S}_{8}$: the summation of binding affinity changes ($\Delta\Delta G$ kcal/mol) by changing RBD residues within 10{\AA } and 8{\AA } to the binding site to glycine; $d$: edge density; $\rho$: degree heterogeneity; $\langle L \rangle$: average path length; $\langle C_b \rangle$: average betweenness centrality; $\langle C_e \rangle$: average eigencentrality; $\langle C_s \rangle$: average subgraph centrality; $\langle M \rangle$: network communicability.
}
\begin{tabular}{@{}|c||r|r|r|r|r|r|}  \hline
\multicolumn{1}{|c||}{Molecule} & \multicolumn{1}{c|}{80R} & \multicolumn{1}{c|}{ACE2} & \multicolumn{1}{c|}{VHH-72} & \multicolumn{1}{c|}{m396} & \multicolumn{1}{c|}{S230} & \multicolumn{1}{c|}{F26G19} \\ \hline
\multicolumn{1}{|c||}{PDB ID} &
\multicolumn{1}{c|}{2GHW} & \multicolumn{1}{c|}{3D0G} & \multicolumn{1}{c|}{6WAQ} & \multicolumn{1}{c|}{2DD8} & \multicolumn{1}{c|}{6NB6} & \multicolumn{1}{c|}{3BGF} \\ \hline \hline
\multicolumn{1}{|c||}{$\Delta G$}            & -17.3 & -11.5 & -9.7 & -9.4 & -7.7 & -6.7\\ \hline
\multicolumn{1}{|c||}{$\text{R}_{10}$}       & 378 & 270 & 255 & 304 & 195 & 254 \\ \hline
\multicolumn{1}{|c||}{$\text{R}_{8}$}        & 157 & 101 & 103 & 119 & 72 & 101 \\ \hline
\multicolumn{1}{|c||}{$\text{S}_{10}$}       & 20.35 & 13.05 & 13.37 & 10.79 & 11.48 & 14.01 \\ \hline
\multicolumn{1}{|c||}{$\text{S}_{8}$}        & 8.92 & 1.56 & 7.49 & 6.47 & 8.09 & 7.84 \\ \hline
\multicolumn{1}{|c||}{$d$}                   & 0.070 & 0.070 & 0.074 & 0.073 & 0.078 & 0.072 \\ \hline
\multicolumn{1}{|c||}{$\rho$}                & 0.0206 & 0.0187 & 0.0193 & 0.0186 & 0.0197 & 0.0190 \\ \hline
\multicolumn{1}{|c||}{$\langle L \rangle$}   & 13.35 & 12.91 & 13.46 & 12.90 & 12.96 & 12.63 \\ \hline
\multicolumn{1}{|c||}{$\langle C_b \rangle$} & 0.0120 & 0.0108 & 0.0109 & 0.0113 & 0.0119 & 0.0113 \\ \hline
\multicolumn{1}{|c||}{$\langle C_e \rangle$} & 0.053 & 0.053 & 0.053 & 0.054 & 0.054 & 0.056 \\ \hline
\multicolumn{1}{|c||}{$\langle C_s \rangle$} & 2693776 & 1418662 & 1607217 & 2383597 & 3167175 & 1897873 \\ \hline
\multicolumn{1}{|c||}{$\langle M \rangle$}   & 1446506 & 730809 & 915061 & 1311299 & 1714397 & 1039376 \\ \hline 
\multicolumn{1}{|c||}{$\langle \Theta \rangle$} & 0.0206 & 0.0187 & 0.0192 & 0.0186 & 0.0196 & 0.0190  \\ \hline
\end{tabular}
\label{tab:SARS1}
\end{table}

\subsection{Prediction of binding affinity changes following mutations}

In this section, we predict the binding affinities of SARS-CoV antibodies when they are applied to SARS-CoV-2 neutralization. Specifically, we compute the binding affinity changes following the mutations from SARS-CoV RBD to SARS-CoV-2 RBD.  These changes can be very significant. Study shows that a single mutation (V367F) can lead to a 10-fold increase in IC$_{50}$ for a particular antibody \cite{rogers2020rapid}. 

\begin{figure}
	\begin{center}
		\includegraphics[width=0.6\linewidth]{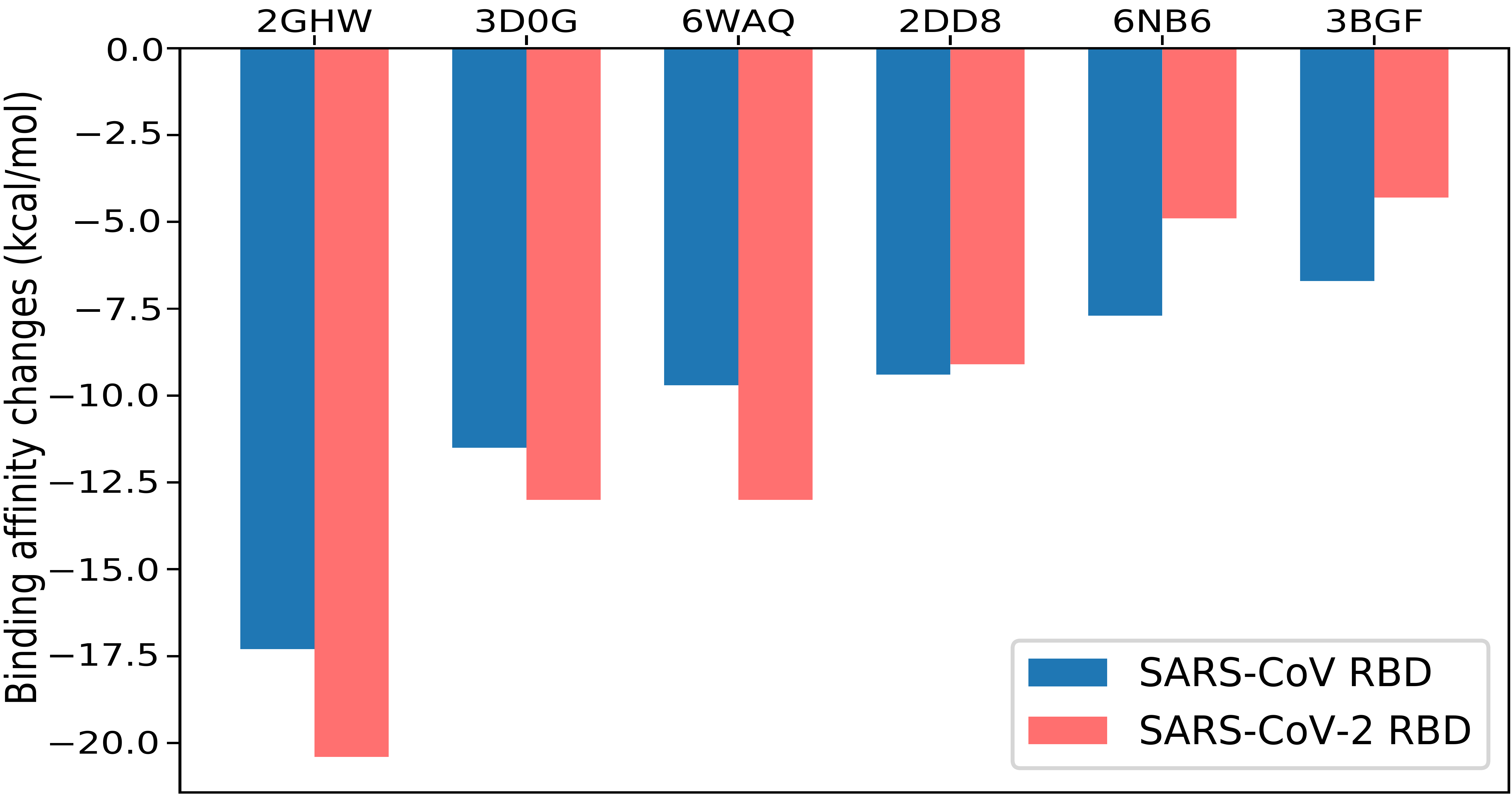}
	\end{center}
	\caption{An illustration of the binding affinities of  antibodies with SARS-CoV and SARS-CoV-2 RBD. Molecular names of these antibodies are 80R (PDB: 2GHW), ACE2 (3D0G), VHH-72 (6WAQ), m396 (2DD8), S230 (6NB6), and F26G19 (3BGF).}
	\label{fig:SARS12}
\end{figure}

In Figure~\ref{fig:SARS12}, the blue bars are predicted binding affinities of each SARS-CoV-2 complex, and the red bars are predicted binding affinities of each molecule with SARS-CoV-2 RBD, which is calculated by accumulating  binding affinities of single mutations from SARS-CoV RBD to SARS-CoV-2 RBD and adding the sum to the binding affinities of SARS-CoV complexes. Obviously, molecule 80R (PDB: 2GHW) has the largest binding energy change in SARS-CoV ranking as well as in SARS-CoV-2 ranking among these SARS-CoV antibodies. Molecule VHH-72 (6WAQ) had a smaller binding affinity than ACE2(3D0G) does for SARS-CoV RBD while an equivalent binding affinity for SARS-CoV-2 RBD. Molecules m396, S230, and F26G19 have weaker binding affinities after modifying from SARS-CoV RBD to  SARS-CoV-2 RBD. Finally, the binding affinity of SARS-CoV RBD with ACE2 following mutations to SARS-CoV-2 is slightly higher than the binding affinity of SARS-CoV RBD with ACE2, indicating that SARS-CoV-2 is more infectious than SARS-CoV. This is consistent with experimental reports     \cite{walls2020structure,wrapp2020cryo}. 

\begin{figure}
	\begin{center}
		\includegraphics[width=0.8\linewidth]{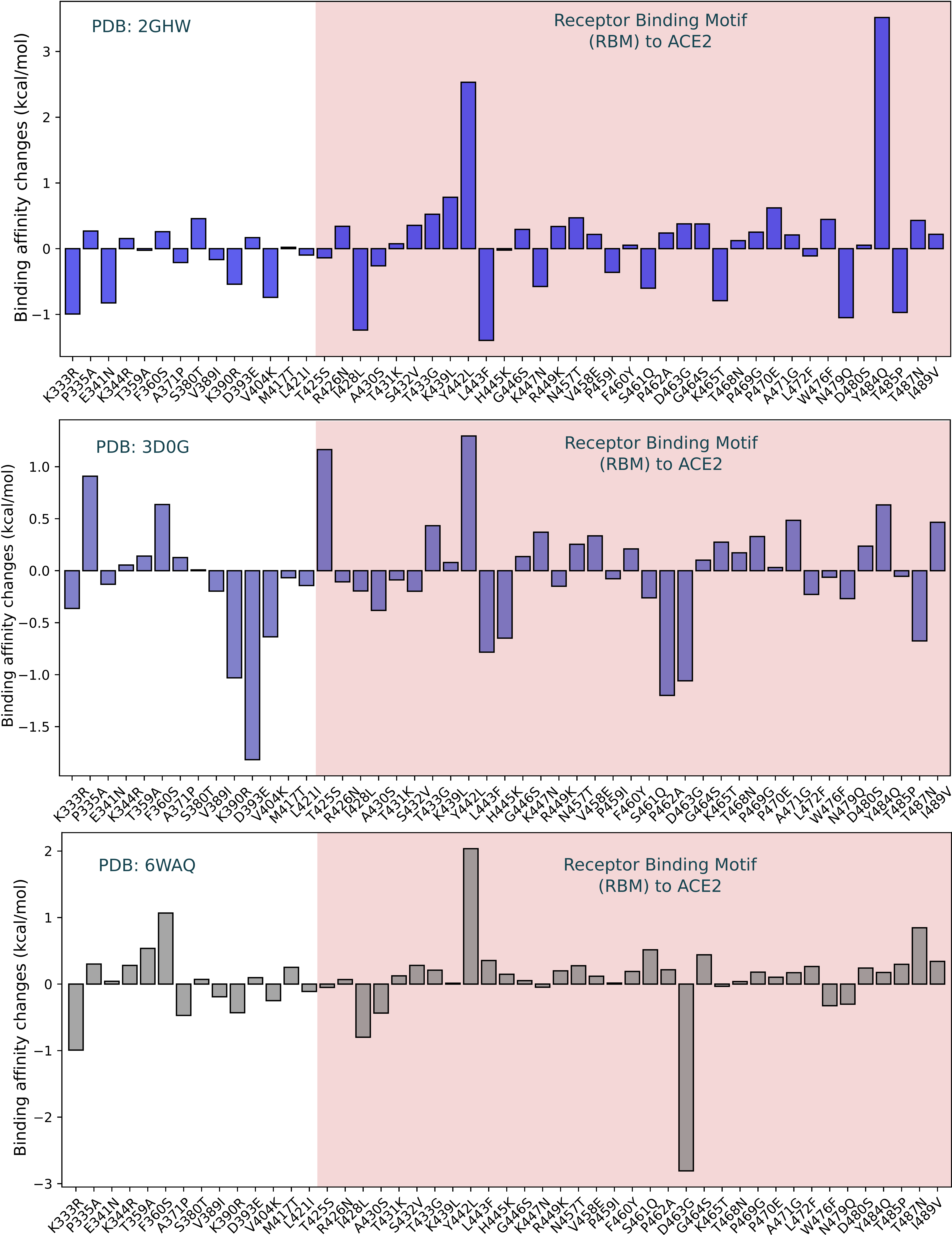}
	\end{center}
	\caption{Overall binding affinity changes on the S protein receptor-binding domain (RBD) from SARS-CoV to SARS-CoV-2 of molecules 80R, ACE2, and VHH-72. The x-axis records the wild type to the mutate type at the specific residue position. The pink color region marks the receptor-binding motif (RBM) corresponding to ACE2. The height of each bar indicates the predicted binding affinity changes. A positive change indicates a strengthening in binding. }
	\label{fig:SARS1_to_21}
\end{figure}

\begin{figure}
	\begin{center}
		\includegraphics[width=0.8\linewidth]{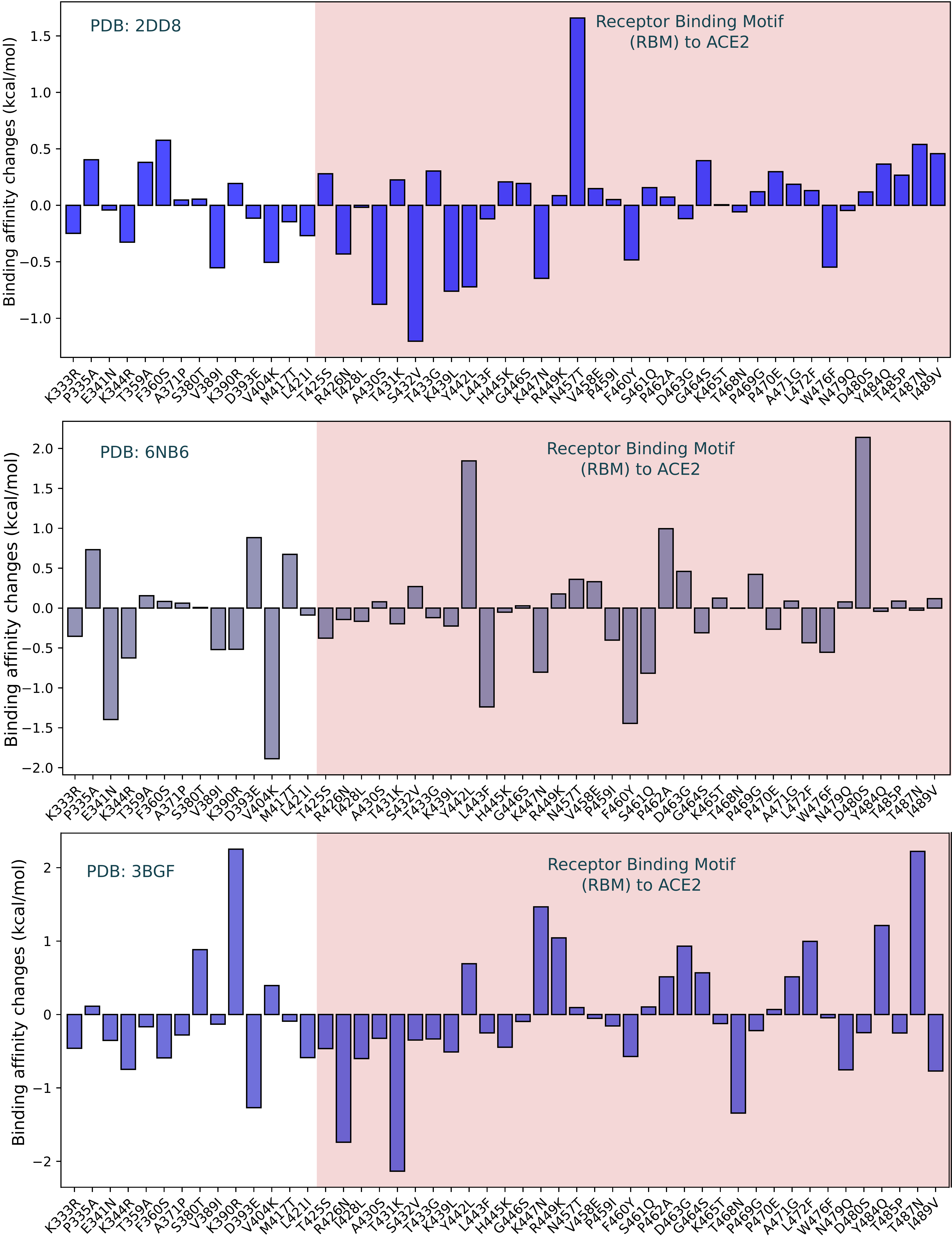}
	\end{center}
	\caption{Overall binding affinity changes on the S protein receptor-binding domain (RBD) from SARS-CoV to SARS-CoV-2 of molecules m396, S320, and F26G19. The x-axis records the wild type to the mutate type at the specific residue position. The pink color region marks  the receptor-binding motif (RBM) corresponding to ACE2. The height of each bar indicates the predicted binding affinity changes. A positive change indicates a strengthening in binding. }
	\label{fig:SARS1_to_22}
\end{figure}

Figures~\ref{fig:SARS1_to_21} and \ref{fig:SARS1_to_22} show the binding affinity changes on individual  mutations on the SARS-CoV S protein RBD, where more precise trends can be observed. In Figure~\ref{fig:SARS1_to_21}, molecule, 80R, has a similar trend to ACE2 which shares the most receipting domain. In Figure~\ref{fig:SARS1_to_22}, most of the binding affinity changes following mutations in receptor binding motif (RBM) of ACE2 are negative, which indicates stronger binding affinities with SARS-CoV RBD.

\subsection{Network analysis of antibody-antigen complexes}
Various network models have been employed to analyze the structure and function of SRAS-CoV and SASR-CoV-2 main protease  \cite{estrada2020topological}. In this work, we utilize network models to illustrate interactions between the binding complexes of the RBD of  SARS-CoV or SRAS-CoV-2 and antibodies or ACE2.  

\begin{figure}
\begin{center}
\includegraphics[width=0.8\linewidth]{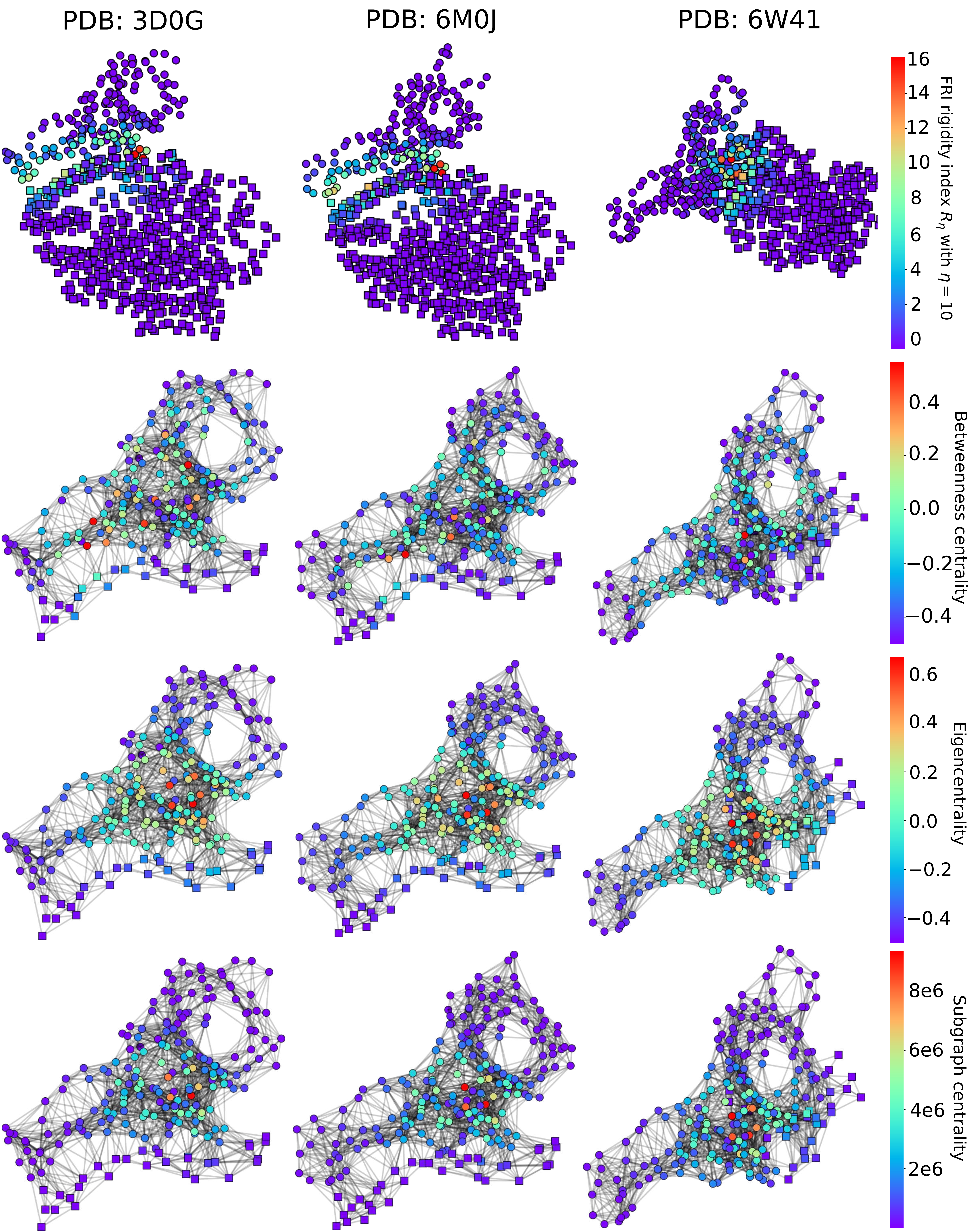}
\end{center}
\caption{C$_\alpha$ network analysis of three antibody-antigen  complexes. Circle markers are for antigen (spike protein RBD) and cube markers are for antibody or ACE2. Columns list  complexes 3D0G, 6M0J, and 6W41. Rows represent FRI rigidity index, betweenness centrality, eigencentrality, and subgraph centrality.}
\label{fig:Combine1}
\end{figure}

In the microscopy of each single residue, their performances on network models reveal the similarities and differences between complexes. In Figure~\ref{fig:Combine1}, SARS-CoV RBD with ACE2 (PDB: 3D0G), SARS-CoV-2 RBD with ACE2 (PDB: 6M0J), and SARS-CoV-2 RBD with CR3022 (PDB: 6W41) are listed and aligned where 6W41 has the strongest predicted binding affinity and the largest deviation to 6M0J as shown in  Tables~\ref{tab:SARS2} and  \ref{tab:SARS1}.  The C$_\alpha$ atoms from the RBD are marked as a circle, and atoms from ACE2 or CR3022 are marked as a cube. In the first row of Figure~\ref{fig:Combine1}, it is interesting to observe all three complexes having similar domains that have high rigidity index values. In the comparison of betweenness centralities of three structures, though it is shown clearly that 3D0G has many large values, it has the lowest average betweenness centrality among the structures as shown in Tables~\ref{tab:SARS2} and  \ref{tab:SARS1}. Analogy to the rigidity index, all three complexes have the same regions of the individual eigencentrality values and subgraph centrality values. Overall, the differences between 3D0G and 6M0J are quite close in their network analysis. However, the betweenness centrality reflects their difference such that a higher average value would indicate  a stronger binding affinity. Moreover, all three complexes coincidentally have similar regions of high values for  network descriptors, which suggests that this region would play a key role in protein-protein interactions.

\section{Clinical trials of COVID-19 antibody therapeutic candidates}

Table \ref{tab:clinical_trials} summarizes the currently ongoing clinical trials of COVID-19 antibody therapeutic candidates in the United States, China  and Europe. These data are collected from the National Institutes of Health (NIH) \url{(https://www.nih.gov/coronavirus)}, the European Medicines Agency (EMA) \url{(https://www.clinicaltrialsregister.eu/ctr-search/search?query=covid-19)}, and the media's reports.

Notably, most of the current COVID-19 antibody therapeutic candidates in clinical trials are aimed at other targets rather than the S protein. These antibodies were developed for other diseases initially and now repurposed to treat COVID-19, which could alleviate some COVID-19 symptoms such as cytokine storm and inflammation instead of killing the viruses. 

Nonetheless, two  antibody candidates under clinical trials are targeting at S protein and block the SARS-CoV-2 entry into human cells. One is LY3819253 developed by Eli Lilly and Company in the United States, which is in phase II clinical trial and already highlighted in TheScientist \url{(https://www.the-scientist.com/news-opinion/first-antibody-trial-launched-in-COVID-19-patients--67604)}. The other is JS016 performed by Junshi Biosciences in China \cite{shi2020human}, which is currently in phase I clinical trial.
 
\begin{center}	
	\begin{longtable}{|C{2.5cm}|C{3cm}|C{1.75cm}|C{4cm}|c|c|}
		\caption{The summary of current on-going clinical trials of Covid-19 antibody therapeutic candidates.}{\footnotemark}
		\label{tab:clinical_trials} \\
		\hline
		Antibody ID & Manufacturer & Target & Trial location & Trial & Start \\ 
		&&&&phase&date \\ \hline
		
		Lanadelumab$^a$  & Shire & pKal & Radboud University Medical Center, Nijmegen, Netherlands  & 4 & / \\ 
		\hline
		
		Octagam$^a$  & Pfizer & Antibody mixture& Sharp Memorial Hospital, San Diego, California, United States & 4 & 4.28.2020 \\ 	
		\hline
		
		Sarilumab$^{a,b}$ & Regeneron Pharmaceuticals and Sanofi & IL-6  & Assistance Publique - Hôpitaux de Paris, Paris, France; & 3 & 3.25.2020      \\ 
		&&&VA Boston Healthcare System,Boston, Massachusetts, United States  &2 & 4.10.2020 \\		
		\hline
		
		Sirukumab$^{a,b}$ & Janssen Biotech & IL-6 & Sanofi-aventis Recherche et Développement, Chilly-Mazarin, France;& 3 & 3.26.2020 \\
		&&& Loyola University Medical Center,Maywood, Illinois, United States  &2 & 4.24.2020 \\		
		\hline
				
		Canakinumab$^{a,b}$  & Novartis & Interleukin-1$\beta$ & Novartis Investigative Site, Glendale, California,United States; & 3 & 4.30.2020  \\ 
		&&&Novartis Pharma GmbH, Nürnberg, Germany &3 & 4.29.2020 \\		
		\hline
				
		IFX-1$^b$ & InflaRx & C5a & InflaRx GmbH, Jena, Germany & 3 & 3.29.2020 \\ \hline
		
		Lenzilumab$^a$ & Humanigen & GM-CSF & Mayo Clinic, Phoenix, Arizona,United States & 3 & 4.30.2020 \\ 	
		\hline
		
		Mylotarg$^b$  & Celltech and Wyeth & CD33 & UK Research and Innovation,United Kingdom  & 3 & 5.5.2020 \\ 
		\hline
		
		Ravulizumab$^b$ & Alexion & C5 & Alexion Europe SAS, & 3 & 5.7.2020 \\
		&Pharmaceuticals && Levallois-Perret, France && \\
		 \hline
		 
		Tocilizumab$^{a,b}$ & Roche & IL-6 & Queen's Medical Center, Honolulu,Hawaii,United States;  & 3 & 6.1.2020    \\ 
		&&&F. Hoffmann-La Roche Ltd, Basel, Switzerland&3 & 4.6.2020\\
		\hline
		
		Avdoralimab$^b$ & Innate Pharma & C5a & Assistance Publique Hopilaux De Marseille, Marseille, France& 2 & 4.23.2020 \\ 
		\hline
		
		Bevacizumab$^b$ & Genentech & VEGF-A & Fundación para la Investigación Biomédica de Córdoba, Córdoba, Spain & 2 & 4.24.2020 \\ 
		\hline
				
		CERC 002$^a$ & Cerecor & LIGHT & Cape Fear Valley Medical Center, Fayetteville, North Carolina,United States& 2 & 6.9.2020                \\ 
		\hline
		
		Clazakizumab$^a$ & Bristol Myers Squibb and Alder Biopharmaceuticals& IL-6 & Cedars-Sinai Medical Center,Los Angeles, California, United States & 2 & 4.24.2020 \\ 		
		\hline
		
		Gimsilumab$^a$ & Eisai Inc & GM-CSF & UCLA Ronald Reagan Medical Center, Los Angeles, California,United States& 2 & 4.12.2020 \\ 
		\hline
		
		IC14$^a$ & Scripps Research & CD14 & University of Washington, Seattle,Washington, United States & 2 & 7.2020 \\  \hline
		
		Infliximab$^a$ & Janssen Biotech & TNF$\alpha$ & Tufts Medical Center,Boston, Massachusetts, United States& 2 & 6.1.2020 \\ 
		\hline
		
		Leronlimab$^a$ & CytoDyn & CCR5 & University of California, Los Angeles, California, United States& 2 & 4.1.2020 \\ 
		\hline
		
		LY3127804$^a$  & Eli Lilly and Company& Ang2 & NorthShore University HealthSystem, Evanston, Illinois, United States & 2 & 4.20.2020 \\ 
		\hline
		
		LY3819253$^a$  & Eli Lilly and Company& S protein & Cedars Sinai Medical Center, Los Angeles, California, United States & 2 & 6.13.2020 \\ 
		\hline	
		
		Mavrilimumab$^a$ & MedImmune & GM-CSF & Cleveland Clinic Health System,Cleveland, Ohio, United States & 2 & 5.20.2020 \\ 
		\hline

		MSTT1041A$^a$ & Genentech & ST2 & eStudySite-Chula Vista-PPDS,Chula Vista, California, United States & 2 & 6.2.2020 \\	
	    \hline
		
		Nivolumab$^b$ & Bristol-Myers Squibb & PD-1 & Centre Léon Bérard, Léon, France & 2 & 4.1.2020 \\ \hline
		
		Otilimab$^{a,b}$  & MorphoSys & GM-CSF & GSK Investigational Site, Saint Louis Park, Minnesota, United States; & 2 & 5.28.2020  \\
		&&&GlaxoSmithKline Research Development Limited, Brentford, United Kingdom&2&5.20,2020 \\
		\hline		
		
		Siltuximab$^b$ & Eusapharma & IL-6 & Fundació Clínic per a la Recerca Biomèdica, Barcelona, Spain& 2 & 4.7.2020                        \\
		\hline
		
		SNDX-6352$^a$ & Syndax Pharmaceuticals & CSF-1R & HonorHealth,Scottsdale, Arizona, United States  & 2 & 5.30.2020 \\
		\hline 
		
		ARGX-117$^b$  & Argenx & C2 & Argenx BV, Zwijnaarde, Belgium & 1 & 4.21.2020                       \\ \hline
		
		TJ003234$^a$  & / & GM-CSF  & GW Medical Faculty Associates,Washington, District of Columbia,United States  & 1  & 4.11.2020 \\ 
		\hline
		
		JS016$^c$ & Junshi Biosciences & S protein & Huashan Hospital Affiliated to Fudan University, Shanghai, China & 1 & 6.7.2020 \\
		\hline	
	\end{longtable}
\end{center}
\footnotetext{List of links of antibodies in \autoref{tab:clinical_trials}: \\
$^a$\url{https://clinicaltrials.gov/ct2/results?recrs=ab\&cond=covid-19\&term=\&cntry=US\&state=\&city=\&dist=}\\
$^b$\url{https://www.clinicaltrialsregister.eu/ctr-search/search?query=covid-19} \\
$^c$\url{https://www.globenewswire.com/news-release/2020/06/07/2044620/0/en/Junshi-Biosciences-Announces-Dosing-of-First-Healthy-Volunteer-in-Phase-I-Clinical-Study-of-SARS-CoV-2-Neutralizing-Antibody-JS016-in-China.html}}

\section{Material and methods}

\subsection{Sequences and structures}
All the sequences and 3D structures are downloaded from Protein Data Bank (\url{https://www.rcsb.org}): the sequences are from the FASTA files, the 3D structures are from the pdb files.

The 3D alignments as well as the graphs are created by using PyMOL \cite{delano2002pymol}. The 2D sequence alignment are calculated by clustalw (\url{https://www.genome.jp/tools-bin/clustalw}) \cite{thompson2003multiple}, the 2D alignment graphs are generated by Jalview \cite{waterhouse2009jalview}. 

\subsection{TopNetTree model for protein-protein interaction (PPI) binding affinity changes upon mutation}
\begin{figure}
\begin{center}
\includegraphics[width=0.9\linewidth]{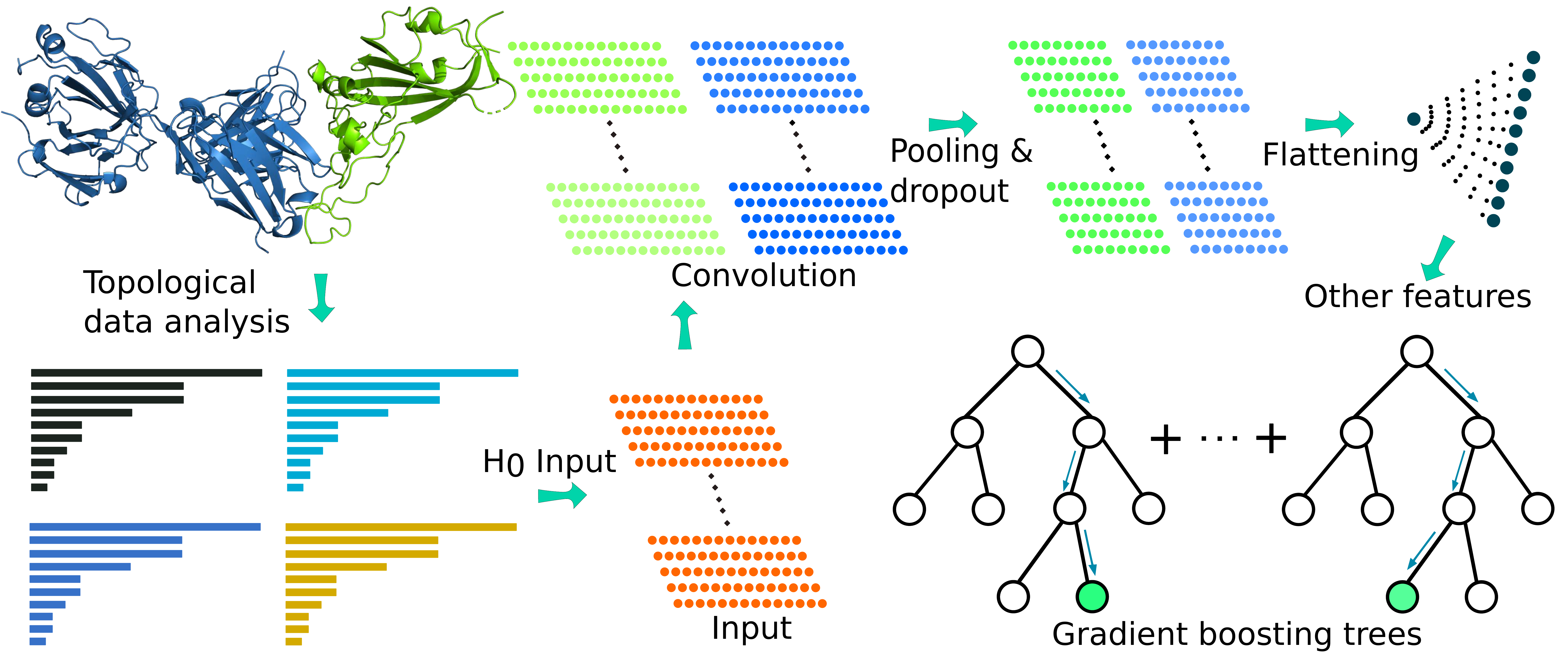}
\end{center}
\caption{An illustration of the TopNetTree model \cite{wang2020topology}. Protein structure shown in the plot is SARS-CoV-2 spike receptor-binding domain bound with antibody (PDB 7BZ5). Here, $H_0$ are the 0-dimensional topological  input features for machine learning model.}
\label{fig:NetTree}
\end{figure}

In this section, the topology-based network tree (TopNetTree) is presented, which predicts binding affinity changes following mutation $\Delta\Delta G$ for PPIs  \cite{wang2020topology}. This method is based on structures regarded as topological features and supervised machine learning model, gradient boosting tree (GBT), and convolution neural network (CNN). In Fig.~\ref{fig:NetTree}, the train and predicting process is elucidated that two major modules are applied the topology-based feature generation and a CNN-assisted GBT model. In feature generation, the element- and site-specific persistent homology is the key mathematical technique that can simplify the structural complexity of protein-protein complexes and translate the biological information into topological invariants. The first step of the TopNetTree process uses CNN as an assistant model to manipulate the topological features. Assembling CNN-pretrained features and other features, the last step of the GBT model predicts PPIs binding affinity changes. Other features such as chemical and physical information that have not been absorbed into topological features can improve the proposed model's predicting ability. More details are referred to in the literature \cite{wang2020topology}.

\subsubsection{Topology-based feature generation of PPIs}
Persistence homology is the key mathematical theory behind the topology-based feature generation. As a subtopic of algebraic topology, persistence homology is built upon simplicial complex and filtration on discrete datasets under various settings. For example, the set of atoms in protein-protein interactions forms the discrete dataset. When building the constructions, a variety of simplicial complex built on point clouds such that Vietoris-Rips (VR) complex and alpha complex are widely used \cite{edelsbrunner2000topological} which are applied in our approach. After a simplicial complex set up, the topological invariants of the point-cloud dataset can be identified and are enumerated by counting the numbers referred to as Betti-0 ($H_0$), Betti-1 ($H_1$), and Betti-2 ($H_2$) for components, rings, and cavities of dataset, respectively. Obviously, the complexity protein-protein structure is simplified as its geometric and topological characteristics for data features, while redundant and uninformative features or calculations are fully abandoned. Moreover, making filtration on simplicial complex turns the 3D point-cloud dataset of PPIs into topological barcodes, which record the ``birth'' and ``death'' of each topological invariants. The topological features simplify the PPI-complex in many directions. However, it is also essential to have better construction to reflect different biological or chemical properties. Various subsets are defined as following for PPI complex constructions.
\begin{enumerate}%[ {(}1{)} ]
	\item $\mathcal{A}_\text{m}$: atoms of the mutation sites.
	\item $\mathcal{A}_\text{mn}(r)$: atoms in the neighbourhood of the mutation site within a cut-off distance $r$.
	\item $\mathcal{A}_\text{Ab}(r)$: antibody atoms within $r$ of the binding site.
	\item $\mathcal{A}_\text{Ag}(r)$: antigen atoms within $r$ of the binding site.
	\item $\mathcal{A}_\text{ele}(\text{E})$: atoms in the system that has atoms of element type E.
\end{enumerate}
Therefore, the distance matrix is defined based on atom sets such that it excludes the interactions in the same set. For interactions between atoms $a_i$ and $a_j$ in set $\mathcal{A}$ and/or set $\mathcal{B}$, the modified distance is defined as
\begin{equation}
D_{\text{mod}}(a_i, a_j) =
\begin{cases}
\infty, \text{ if } a_i, a_j\in\mathcal{A}\text{, or }a_i, a_j\in \mathcal{B}, \\
D_e(a_i, a_j), \text{ if } a_i\in\mathcal{A} \text{ and }a_j\in \mathcal{B},
\end{cases}
\label{eq:modified_equation}
\end{equation}
where $D_e(a_i, a_j)$ is the Euclidian distance between $a_i$ and $a_j$. Next, the persistence homology can be constructed element- and site-specific.

Given atomic coordinates, their topological analysis and properties can be carried out via simplices and simplicial complexes. A set of $k\!+\!1$ affinely independent points, $v_0$, $v_1$, $v_2$, $...$, $v_k$ in $\mathbb{R}^n$ is a $k$-simplex denoted $\sigma_i$, such that a 0-, 1-, 2-, or 3-simplex in geometry representation is a vertex, an edge, a triangle, or a tetrahedron, respectively. The finite collection of simplex is a simplicial complex $K=\{\sigma_i\}$, which is true if a subset (also called as face) $\tau$ of a $k$-simplex $\sigma_i$ of $K$ is also in $K$, $\tau \subseteq \sigma_i$ and $\sigma_i\in K$ imply $\tau \in K$ and the non-empty intersection of any two simplices in $K$ is a face of both. Furthermore, a $k$-chain is a finite formal sum of all simplices in $K$, $\sum_{i}\alpha_i\sigma^k_i$, where $\alpha_i$ is coefficients in $\mathbb{Z}_p$ and $p$ is a chosen prime number. The set of all $k$-chains of the simplicial complex $K$ equipped with an algebraic field forms an abelian gourp $C_k(K, \mathbb{Z}_p)$. 

A boundary operator $\partial_k: C_k\!\rightarrow\!C_{k-1}$ for a $k$-simplex $\sigma^k$ is homomorphism defined as \[\partial_k \sigma^k = \sum^{k}_{i=0} (-1)^i\{ v_0, v_1, \cdots, \hat{v_i}, \cdots, v_k \},\] where $\{v_0, v_1, \cdots ,\hat{v_i}, \cdots, v_k\}$ is a $(k\!-\!1)$-simplex excluding $v_i$ from the vertex set. An important property of boundary operator, $\partial_{k-1}\partial_k= \emptyset$, follows from that boundaries are boundaryless. Moreover the kernel of boundary operator is $Z_k={\rm ker} \partial_k=\{c\in C_k \mid \partial_k c=\emptyset\}$, whose elements are called $k$-cycles; and the $k$th boundary group is the image of $\partial_{k+1}$ denoted as $B_k={\rm im} ~\partial_{k+1}= \{ \partial_{k+1} c \mid c\in C_{k+1}\}$. The algebraic construction to connect a sequence of complexes by boundary maps is called a chain complex
\[
\cdots \stackrel{\partial_{i+1}}\longrightarrow C_i(X) \stackrel{\partial_i}\longrightarrow C_{i-1}(X) \stackrel{\partial_{i-1}}\longrightarrow \cdots \stackrel{\partial_2} \longrightarrow C_{1}(X) \stackrel{\partial_{1}}\longrightarrow C_0(X) \stackrel{\partial_0} \longrightarrow 0
\]
and the $k$th homology group is the quotient group defined by $H_k = Z_k / B_k$. Obviously, boundary operators imply $B_k\subseteq Z_k \subseteq C_k$. The Betti numbers are defined by the number of basis in $k$th homology group $H_k$ which counts $k$-dimensional holes. For example, Betti 0, $\beta_0\!=\!{\rm rank}(H_0)$ reflects the number of connected components; Betti 1, $\beta_1\!=\!{\rm rank}(H_1)$ reflects the number of loops; and Betti 2, $\beta_2\!=\!{\rm rank}(H_2)$ reveals the number of voids or cavities. Together, the set of Betti numbers $\{\beta_0,\beta_1,\beta_2,\cdots \}$ indicates the intrinsic topological property of a system. Computational, boundary operators directly work on the distance matrices generated on different atom groups, and the Betti number can be calculated by counting the number of zero eigenvalues of corresponding boundary operators.

It is interested in the evolution of a simplicial complex and to track topological characteristics varying as the simplicial complex changes. In persistent homology, a filtration of a topology space $K$ is a nested sequence of subspaces $\{K^t\}_{t=0,...,m}$ of $K$ such that $\emptyset = K^0 \subseteq K^1 \subseteq K^2 \subseteq \cdots \subseteq K^m = K$. Considering the complex group on this sequence, we can have a sequence of chain complexes by homomorphisms: $C_*(K^0) \to C_*(K^1) \to \cdots \to C_*(K^m)$ and a corresponding homology sequence: $H_*(K^0) \to H_*(K^1) \to \cdots \to H_*(K^m)$. The $p$-persistent $k$th homology group of $K^t$ is defined as $H_k^{t,p} = Z^t_k/(B_k^{t+p}\bigcap Z^t_k)$, where $B_k^{t+p} = {\rm im} \partial_{k+1}(K^{t+p})$. Hence, the homology group reveals that the homology classes of $K^t$ persist until $K^{t+p}$. In the filtration process, the persistent homology barcodes recording the ``birth'' and ``death'' of topological invariants can be generated along the spacial changing of radius on point-cloud dataset. The machine learning feature vectors, as consequences, can be constructed from theses sets of filtration barcode intervals.

The filtration parameter interval is discretized into bins, which can model the behavior of barcodes in each bin \cite{cang2018representability}. Thus, these bins are packaged as features for advanced machine learning algorithms directly. Then the number of persistence intervals is counted for each bin in order to record birth events and death events. Three feature vectors ($H_0$, $H_1$, and $H_2$) are generated for each topological barcode for the machine learning method. Betti-0 ($H_0$) barcode is obtained from the VR filtration and Betti-1 ($H_1$), and Betti-2 ($H_2$) barcode is obtained from alpha complex filtration, where Betti-1 and Betti-2 are sparser and more stable than Betti-0 barcodes. Thus, Betti-0 barcode is incorporated into CNN models, and Betti-1 and Betti-2 barcodes are for GBT training. Intuitively, features generated by binned barcode vectorization can reflect the structure of the protein-protein complex and its biological and chemical properties, such as the strength of atom bonds, van der Waals interactions. Meanwhile, taking the statistics of bar lengths, birth values, and death values, such as maximum, minimum, mean, etc., can be set as features for the machine learning process.

%\subsubsection{Other features generation of PPIs}

\subsubsection{Machine learning models}
To predict the binding affinity changes following mutations for PPIs is very challenging due to the complex dataset and different 3D structures. To overcome this challenge, a hybrid machine learning algorithm which integrates a CNN and GBT to predict the binding affinity changes. The vectorized $H_0$ barcode feature is converted into concise features by the CNN method. Then combine CNN-trained features and the rest features as the full feature set to train a GBT module for a robust predictor with effective control of overfitting.

CNN is considered as the most successful architectures as a class of deep neural networks. CNN is a regularized case of a multilayer connected neural network. Each neuron is connected locally to the next convolution layer neurons, and the weights are shared in different locations. In TopNetTree, CNN is an intermediate model that applies vectorized $H_0$ features into a higher-level abstract feature for the gradient boosting tree method. Next, the GBT is an ensemble method that builds a powerful module for regression and classification problems as weak learners. The method sums the weak learners to eliminate the overall error based on the assumption that each learner is likely to make different mistakes. According to the current prediction error on the training dataset, the ensemble method is built upon a decision tree structure. GBT with topological features (TopGBT) is relatively robust against hyperparameter tuning and overfitting, and is suitable for a moderate number of features. The current work uses the GBT package provided by scikit-learn (v 0.23.0) \cite{pedregosa2011scikit}. 

Finally, TopNetTree follows the process (Fig.~\ref{fig:NetTree}) that a supervised CNN model is trained for extracting high-level features from $H_0$ barcodes, where the PPI $\Delta\Delta G$ is a label. Then the flatten layer neural outputs of CNN are ranked as their importance in a GBT model. Based on the importance, the whole features consist of an ordered subset of CNN-trained features, high-dimensional topological barcodes, $H_1$ and $H_2$, and auxiliary features for the final GBT model. As for the parameters, an optimal parameter setting with the best result of the 10-fold evaluation is selected from the experiments of different parameter settings.

\subsubsection{Cross-validation of TopNetTree}
\begin{table}[!htb]
	\caption{Ten-fold cross-validation of the TopNetTree on the SKEMPI 2.0 dataset.  }
	\centering
	
	\begin{tabular}{lccc|lccc}
		\toprule
		& $R_p$ & $\tau$ & RMSE (kcal/mol) & & $R_p$ & $\tau$ & RMSE (kcal/mol)\\
		\midrule
		Fold 1 (Train) & 0.981 & 0.884 & 0.366 & Fold 6 (Train) & 0.983 & 0.904 & 0.353 \\
		Fold 1 (Test)  & 0.835 & 0.595 & 1.065 & Fold 6 (Test)  & 0.836 & 0.594 & 1.064 \\
		\midrule
		Fold 2 (Train) & 0.982 & 0.902 & 0.360 & Fold 7 (Train) & 0.983 & 0.904 & 0.356 \\
		Fold 2 (Test)  & 0.839 & 0.600 & 1.061 & Fold 7 (Test)  & 0.838 & 0.594 & 1.060\\
		\midrule
		Fold 3 (Train) & 0.982 & 0.887 & 0.366 & Fold 8 (Train) & 0.979 & 0.878 & 0.392 \\
		Fold 3 (Test)  & 0.837 & 0.595 & 1.068 & Fold 8 (Test)  & 0.840 & 0.596 & 1.061 \\
		\midrule
		Fold 4 (Train) & 0.981 & 0.880 & 0.369 & Fold 9 (Train) & 0.982 & 0.902 & 0.362 \\
		Fold 4 (Test)  & 0.841 & 0.596 & 1.059 & Fold 9 (Test)  & 0.838 & 0.596 & 1.069\\
		\midrule
		Fold 5 (Train) & 0.982 & 0.906 & 0.365 & Fold 10 (Train)& 0.982 & 0.886 & 0.367 \\
		Fold 5 (Test)  & 0.839 & 0.594 & 1.062 & Fold 10 (Test) & 0.835 & 0.596 & 1.064 \\
		\midrule
		Average (Train) & 0.982 & 0.893 & 0.366 & & & &  \\
		Average (Test)  & 0.838 & 0.596 & 1.063 & & & & \\
		\bottomrule
	\end{tabular}
	\label{tab:10-fold}
\end{table}

The proposed TopNetTree method is trained on the SKEMPI 2.0 dataset \cite{jankauskaite2019skempi}, which has 4,169 variants in 319 different complexes. A set ``S8338'' with 8,338 variants was derived from SKEMPI 2.0 dataset by setting the reverse mutation energy changes to the negative values of its original energy changes. To address the reliability of the TopNetTree method, we did the tenfold cross-validation on the SKEMPI 2.0 dataset with the averaged training accuracy, Pearson correlation coefficients $R_\text{p}$, Kendall's $\tau$, and the root mean square error (RMSE), being 0.98, 0.89, and 0.37 kcal/mol. As shown in Table~\ref{tab:10-fold}, these metrics are based on the average of ten ten-fold cross-validations, which indicate TopNetTree is well trained. The performance test of tenfold cross-validation on dataset gives as $R_\text{p}=0.84$, $\tau = 0.60$, and RMSE $=1.06$ kcal/mol, which is of the same level of accuracy as the best in the literature \cite{wang2020topology}. 

\subsection{Graph network analysis}
Graph networks represent interactions between pairs of units in biomolecular systems. The quantify unique characteristics of the networks can be measured for descriptions and comparisons of different networks. Considering the protein-protein interactions as networks, each descriptor evaluates the network properties and measures how proteins connect. For instance, a fixed domain of spike protein RBD and antibodies forms a network, where residues from 320 to 518 on SARS-CoV and residue from 329 to 530 on SARS-CoV-2 are considered in terms of C$_{\alpha}$ atoms. As aforementioned interaction subsets or similar subsets are defined for  C$_\alpha$ of each amino acid as following.
\begin{enumerate}%[ {(}1{)} ]
\item $\mathcal{C}_\text{Ab}(r)$: antibody $\text{C}_\alpha$ atoms within $r$ {\AA } of the binding site, where $r=\infty$ is for all $\text{C}_\alpha$ atoms on  antibody.
\item $\mathcal{C}_\text{Ag}(r)$: antigen $\text{C}_\alpha$ atoms within $r$ {\AA } of the binding site, where $r=\infty$ is for all $\text{C}_\alpha$ atoms on  antigen.
\end{enumerate}
With these definitions, network descriptors are defined below.

{\bf FRI rigidity index} The FRI rigidity index is a great tool to illustrate the elasticity between atoms for molecular interaction prediction \cite{nguyen2016generalized, xia2013multiscale}. The molecular rigidity index is defined as a summation of all the atomic rigidity index $\mu_{\eta,i}$ as
\begin{equation}
\label{eq:rigidity}
R_{\eta} = \sum_{i=1}^{N_{AB}}\mu_{\eta, i}= \sum_{i=1}^{N_{AB}} \sum_{j=1}^{N_{AG}}  e^{-\big(\frac{\|\textbf{r}_i-\textbf{r}_j\|}{\eta}\big)^2},
\end{equation}
where $\textbf{r}_i$ are atom positions, $N_{AB}$ and $N_{AG}$ are the numbers of atoms of antibody and antigen, respectively, and $r=\infty$ for all $\text{C}_\alpha$ atoms such that $\mathcal{C}_\text{Ab}(\infty)$ and $\mathcal{C}_\text{Ag}(\infty)$. TThe molecular rigidity index is used to describe the behavior of the dynamics and elastostatics of the biomolecular elasticity where $\eta$ controls the influence range between atoms. In PPIs, the elasticity between antibody and antigen, especially long range impacts, is studied by calculating the FRI index of the network consisting of  {C}$_\alpha$ atoms. 

{\bf Degree heterogeneity} The degree heterogeneity is an index that evaluates the heterogeneity of a network on different distribution \cite{estrada2010quantifying}. The degree distribution $k_i$ is the number of $i$-th nodes that have $k_i$ connections to other nodes. Therefore, the degree heterogeneity reflects the distributions of a network on different impacts, which is defined as 
\begin{equation}
\rho=\sum_{i=1}^{N_e}\sum_{j=i+1}^{N_e}(k_i^{-1/2}-k_j^{-1/2})^2.
\end{equation}
Here, $N_e$ represents for the number of edges. In our case, we study two networks consisting of all  {C}$_\alpha$ atoms in $\mathcal{C}_\text{Ag}(\infty)$, that one network consists of $\text{C}_\alpha$ atoms from SARS-CoV RBD and SARS-CoV-2 RBD. The degree heterogeneity illustrate the impacts of ACE2  or antibodies to these networks.

The rest descriptors are build on the network consist of  {C}$_\alpha$ atoms from $\mathcal{C}_\text{Ag}(\infty)$ and $\mathcal{C}_\text{Ab}(10)$.

{\bf Edge density} The edge density is defined as 
\begin{equation}
d=\frac{2N_e}{N_v(N_v-1)},
\end{equation}
where $N_e$ is the number of edges and $N_v$ is the number of vertices for C$_\alpha$ atoms from $\mathcal{C}_\text{Ag}(\infty)$ and $\mathcal{C}_\text{Ab}(10)$. The edge density is also called the average degree centrality. For a complete network in which each every pair of network vertices is connected, the edge density is equal to one. A non-complete network has an edge density smaller than one. With the same number of residues in RBD for each PPI, a higher edge density stands for a firmly connection between RBD and ACE2 or antibodies.

{\bf Average path length} The characteristic path length studies the typical separation between two vertices in the network. It was used to study infectious diseases spread in so called ``small-world'' networks \cite{watts1998collective}. The shortest path distance $d(i,j)$ is defined as the shortest path between the corresponding pairs of vertex $i$ and $j$. In protein-protein interactions, the path length between two atoms reflects how ACE2 or antibodies connect to RBD. The average path length is defined as 
\begin{equation}
\label{eq:ave_path_length}
\langle L\rangle = \frac{1}{N_v(N_v-1)} \sum_{i=1}^{N_v}\sum_{j=i+1}^{N_v} d(i,j),
\end{equation}
for $\text{C}_\alpha$ atoms from $\mathcal{C}_\text{Ag}(\infty)$ and $\mathcal{C}_\text{Ab}(10)$. Here, $N_v$ represents for the number of vertices. 

{\bf Average betweenness centrality} The concept of betweenness centrality illustrates communications in a network \cite{freeman1978centrality}. The betweenness centrality of a vertex $v_k$ is given as 
\begin{equation}
C_b(v_k) = \sum_{i=1}^{N_v}\sum_{j=i+1}^{N_v} g_{ij}(v_k)/g_{ij}
\end{equation}
and the average betweenness centrality is given as
\begin{equation}
\langle C_b \rangle = \frac{1}{N_v}\sum_{k=1}^{N_v} C_b(v_k),
\end{equation}
where $g_{ij}(v_k)$ is defined as the number of geodesics linking between vertex $v_i$ and $v_j$ that passes $v_k$, and $g_{ij}$ considers all the paths between $v_i$ and $v_j$. And $N_v$ means the number of vertices. 

{\bf Average eigencentrality} The eigenvector centrality are the elements of the eigenvector $V_\text{max}$ with respect to the largest eigenvalue of the adjacency matrix $A$ \cite{bonacich1987power}. It describes the probability of starting and returning at the same point for infinite length walks. Thus the average eigenvector centrality is, 
\begin{equation}
\langle C_e \rangle = \frac{1}{N_v}\sum_{i=1}^{N_v}e_i,
\end{equation}
where $e_i$ are elements of $V_\text{max}$. It stands for the average impact spread of vertices beyond its neighborhood for an infinite walk.

{\bf Average subgraph centrality} The following descriptors are built on the exponential of the adjacency matrix, $E = e^A$. The average subgraph centrality is defined as 
\begin{equation}
\langle C_s \rangle = \frac{1}{N_v}\sum_{k=1}^{N_v} E(k,k),
\end{equation} 
which indicates the vertex participating in all subgraphs of the graphs \cite{estrada2005subgraph, estrada2020topological}. Here, $E(k,k)$ means the element located at the $k$-th row and $k$-th column. Subgraph centrality is the summation of weighted closed walks of all lengths starting and ending at same node. The long path length has a small contribution.

{\bf Average communicability} Finally, the last two descriptors are average communicability given as
\begin{equation}
\langle M \rangle = \frac{2}{N_v(N_v-1)}\sum_{i=1}^{N_v}\sum_{j=i+1}^{N_v} E(i,j),
\end{equation}
and average communicability angle given as
\begin{equation}
\langle \Theta \rangle = \frac{2}{N_v(N_v-1)}\sum_{i=1}^{N_v}\sum_{j=i+1}^{N_v} \theta(i,j),
\end{equation}
where $\theta(i,j) = \arccos \Big( \frac{E(i,j)}{\sqrt{E(i,i),E(j,j)}} \Big)$ and $E$ is the exponential of the adjacency matrix. The average communicability measures how much two vertices can communicate by using all the possible paths in the network, where the shorter path has more weight \cite{estrada2008communicability}. The average communicability angle evaluates the efficiency of two vertices passing impacts to each other in the network with all possible paths \cite{ estrada2016communicability,estrada2020topological}. 

\section*{Conclusion}
Currently, developing effective therapies for combating Coronavirus disease 2019 (COVID-19) caused by  severe acute respiratory syndrome coronavirus 2 (SARS-CoV-2) becomes a vital task for human health and the world economy. Although designing new anti-SARS-CoV-2 drugs is of paramount importance, traditional drug discovery might take many years. Effective vaccines typically require more than a year to develop. Therefore, a more efficient strategy in fighting against COVID-19 is to look for antibody therapies, which is a relatively easier technique compared to the development of small-molecular drugs or vaccines. Seeking possible antibody drugs  has attracted increasing attention in recent months. Moreover, complementarity determining regions (CDRs) which located in the tip of the antibody, determine the specificity of antibodies and make the antibody therapies a promising way to fight against COVID-19. We analyze the structure, function, and therapeutic potential of seven SARS-CoV-2 neutralizing antibody candidates that have three-dimensional (3D) structures available in the Protein Data Bank. In a comparative study, we also review five antibody structures associated with SARS-CoV and angiotensin-converting enzyme 2 (ACE2) complexes with both  SARS-CoV-2 and SARS-CoV spike proteins. The multiple order of magnitude discrepancies in reported experimental binding affinities motivates us to carry a systematic computational analysis of the aforementioned fourteen complexes. Using computational topology, machine learning, and wide class network models, we put all of the complexes in an equal footing to evaluate binding and interactions. Additionally, we have predicted binding affinities of five SARS-CoV antibodies when they are used as SARS-COV-2 therapies.  Finally, we summarize all of the currently ongoing clinical antibody trails for SARS-CoV-2, which have many targets, including the S protein. In a nutshell, we provide a review of existing antibody therapies for COVID-19 and introduce  many theoretical models to rank the potency and analyze the properties of antibodies.

\section*{Supporting Material}
Supporting Materials are available for  S1: Network analysis of antibody-antigen complexes and 
S2: Binding affinity changes following mutations.

\section*{Acknowledgment}
This work was supported in part by NIH grant  GM126189,   NSF Grants DMS-1721024,  DMS-1761320, and IIS1900473,  Michigan Economic Development Corporation,  Bristol-Myers Squibb,  and Pfizer.
The authors thank The IBM TJ Watson Research Center, The COVID-19 High Performance Computing Consortium, NVIDIA, and MSU HPCC for computational assistance.  

%\bibliographystyle{abbrv}
%% \bibliographystyle{custom}
%\bibliography{refs}

% \end{multicols}
\end{document}